\documentclass[a4paper,twoside]{article}

\usepackage[utf8]{inputenc}

\usepackage{geometry}
\usepackage{fancyhdr}
\usepackage[sorting=none]{biblatex}
\usepackage{subcaption}
\usepackage{graphicx}
\usepackage{pslatex}
\usepackage{hyperref}
\usepackage[dvipsnames]{xcolor}
\usepackage[english]{babel}
\usepackage{amsmath}
\usepackage{amsfonts}
\usepackage{amssymb}
\usepackage{biblatex}
\usepackage{graphicx}
\usepackage{bbold}
\usepackage{nicematrix}
\usepackage{arydshln}
\usepackage{extarrows}
\usepackage{makecell}

\usepackage{tikz}
\usetikzlibrary{quantikz2}

\title{Quantum Lattice Boltzmann Method for Multiple Time Steps Without Reinitialization for Linear Advection-Diffusion Problems}
\author{\underline{Aaron Nagel$^{\ast}$} and Johannes Löwe$^{\ast}$\\
\\
$^{\ast}$German Aerospace Center (DLR)\\
Institute of Aerodynamics and Flow Technology\\
Bunsenstrasse 10, D-37073 Goettingen, Germany\\
aaron.nagel@dlr.de $\cdot$  johannes.loewe@dlr.de
}
\date{October 2025}

\begin{document}

\maketitle

\section*{Abstract}

To simulate highly-resolved flow fields, we extend the Quantum Lattice Boltzmann Method (QLBM) to be able to simulate multiple time steps without state extraction or reinitialization. We adjust and extend given QLBM approaches from the literature to completely remove the need to measure or reinitialize the flow field in between the simulation time steps. Therefore, our algorithm does not require to sample the entire flow field at any time. We solve the linear advection-diffusion problem and derive all necessary equations and build the corresponding quantum circuit diagrams, including details on the QLBM blocks and explicitly drawing the circuit gates. We discuss the general decay of a QLBM step and how that effects our algorithm. The new algorithm is verified on 1D and 2D test cases using the \textit{shot} method of IBMs \textit{Qiskit} package. We show excellent agreement and convergence between our QLBM and classical LBM methods.
The conclusion section includes a discussion on the advantages of our algorithm as well as limitations and to what extent it is more efficient.

\section{Introduction}

The effort in developing quantum algorithms has increased rapidly in recent years. But the idea of using quantum particles as operation units in machines, similar as the classical bit in a computer, is not new at all. Richard Feynman as one famous example, has already published first ideas on \textit{Quantum Mechanical Computers} \cite{feynman1986quantum} back in the 1980s, and first important quantum algorithms have been developed in 1990s, the well known \textit{Shor's} \cite{shor1994algorithms} and \textit{Grover's} \cite{grover1996fast} algorithms. But with the advancements in quantum hardware after the year of 2000, also the interest has become more and more present. With the increase of usable qubits, especially the promise of an exponential system size scaling in quantum operation units raised a large interest in scale resolving fluid flow simulations. \\
The first ideas in approaching computational fluid dynamics (CFD) problems with quantum algorithms were proposed by Meyer \cite{meyer1996quantum} and Yepez \cite{yepez2001quantum} with quantum lattice gas approaches. Improved models by the lattice Boltzmann methods (LBM) have been developed further by the pure fluid transport with a collisionless Boltzmann equation \cite{todorova2020quantum,schalkers2022efficient} using a quantum streaming operation. Further work included a quantum collision step by separation of the collision procedure into a sum of unitary operations \cite{budinski2021quantum, childs2012hamiltonian,shinde2025utilizing}. These algorithms represent the first “\textit{fully quantum}" LBM algorithms for a single LBM time step. But such a full LBM routine as presented in the literature requires a full state extraction and reinitialization in between every time step. This again results in a complete loss of its quantum advantage in system size scaling, since a full state extraction scales linear in the grid resolution with the number of shots needed to resolve the flow field. That being said, literature claiming to develop a “\textit{fully quantum}" LBM algorithm has to be judged carefully, since this can mean that a quantum LBM is developed only for a single time step. \\
Additionally to the LBM algorithms of linear advection-diffusion problems, the nonlinearity of the collision operator has been approached using a Carleman linearization within the LBM by Itani et al. \cite{itani2022Carleman,itani2024quantum} and for the Burgers equation by Liu et al. \cite{liu2021efficient}. However, the linearization comes at the cost of temporal instability for large nonlinearities as they commonly occur in typical aerospace problems. Since our paper will not cover nonlinearities, the topic itself has to be addressed in future work. \\
In this paper, we present our algorithm of a quantum lattice Boltzmann method (QLBM) for a linear advection-diffusion equation for multiple time steps which does not require measurement or reinitialization at any time in between the simulation time steps. This is to the best of the authors knowledge the first algorithm of its kind that can perform multiple QLBM time steps without any kind or mid-circuit measurement, state extraction or reinitialization of the quantum state. Our new algorithm can perform all QLBM time steps up to the end of the simulation, without the need to ever \textit{having} to extract the flow field at any time. For verification reasons, we show in our result section that our algorithm reproduces the correct full flow field up to a certain sampling error. But in contrast to other algorithms, our algorithm allows to be used in such a way, that for the flow field properties you are interested in, quantities can be calculated without ever having to extract the flow field, not even at the end of the simulation. This way, a simulation of a body in the flow field could return a scalar value like a lift or drag coefficient value, while the entire fully resolved flow field was never measured. Our algorithm is a first approach to an actually fully quantum algorithm for multiple time steps, even when still limited in the total number of simulation time that is feasible to simulate.\\
The paper is structured as follows: the \textit{Methods} section starts with a short overview of the lattice Boltzmann method (LBM) that is used which will be translated into a quantum lattice Boltzmann method (QLBM) for an advection-diffusion problem. Further subsections describe the necessary quantum state amplitude encoding structure, the quantum collision step that is used, the quantum streaming step that is adopted from the literature, the procedure to calculate the macroscopic values as a quantum state and finally the quantum re-preparation step to make the algorithm work for multiple time steps. The subsections lead through the mathematical derivations and show the implementation of the quantum circuit blocks explicitly by quantum gates. The decay of the quantum state amplitude for multiple time steps is quantified in a final subsection. The following \textit{Verification} section shows 1D and 2D results of an advection-diffusion process and verifies the QLBM to recover the correct LBM solution using the \textit{shot} method of IBMs simulator \textit{Qiskit} \cite{qiskit2024}. For the 2D test cases, deviations and convergence of QLBM to LBM are discussed.
In the final \textit{Conclusion} section, a discussion on the advantage of the algorithm and limitations are included.

\section{Methods}

\subsection{The Advection-Diffusion equation}

In this paper, we focus on solving linear transport equations with the lattice Boltzmann method (LBM), in particular the advection-diffusion equation of a scalar $\Phi$:
\begin{equation}\label{eq:ADE}
    \frac{\partial \Phi}{\partial t}
    +
    u_j\frac{\partial \Phi}{\partial x_j}
    =
    D\frac{\partial^2 \Phi}{\partial x_j^2},
\end{equation}
with a uniform constant flow velocity $u_j$ and a constant diffusion coefficient $D$, using the Einstein's index summation notation over the spacial directions $j \in \{1,2,3\}$. An extension of a spacially variable flow velocity is generally possible with our approach by conditioning the collision operations on different grid locations. This will be addressed  in future work.\\

\subsection{The Lattice Boltzmann Method}

Boltzmann methods generally describe the dynamics of fluids from a statistical perspective of a particle ensemble description of the fluid. This comes with advantages because it allows to describe the fluid dynamic with distinct \textit{streaming} and \textit{collision} steps to determine the change and the transport of the fluid property in time. These two steps are evaluated in two fully separated steps, which then can be developed individually to a quantum algorithm to account for the desired physics modeling. \\
In this work, we will use the lattice Boltzmann method (LBM) using a fairly simple collision description by the Bhatnagar–Gross–Krook (BGK) collision operator to develop our algorithm. \\
\newline
The Boltzmann equation \cite{mohamad2011lattice,kruger2017lattice} is a transport equation that determines the change of a velocity distribution function $f$, which is a probability density function for a local particle ensemble of positions and velocities $f(\mathbf{x}, \mathbf{v}, t)$ in phase space $\mathcal{H}$. The Boltzmann equation describes the change of the local fluid quantity by the advective transport on the left hand side and the change due to the collision by the collision operator $S_\text{coll}(f)$:

\begin{equation}\label{eq:Boltzmann_eq}
    \frac{\partial f}{\partial t} + v_j \frac{\partial f}{\partial x_j} + F_j \frac{\partial f}{\partial v_j}
    =
    S_\text{coll}(f),
\end{equation}
which implies the Einstein's index summation notation over the spacial directions $j \in \{1,2,3\}$ and where $F_j$ can be some additional external acceleration. \\
We assume no external forces $F_j = 0$ and use a collision relaxation by the BGK operator \cite{mohamad2011lattice}
\begin{equation}\label{eq:S_BGK}
    S_\text{BGK}(f)
    =
    -\frac{1}{\tau}\left(
        f - f^\text{eq}
    \right),
\end{equation}
where $\tau$ is the relaxation time towards a local equilibrium $f^\text{eq}$, which is determined by the diffusion coefficient $D$. In further simplifications, we will choose a fixed relaxation time and model the diffusion by the weight parameters of the LBM. \\
The position space is discretized on a grid and the velocity is discretized to
a set of vectors $\mathbf{e}_i$ that span the velocity space, which are not necessarily linear independent vectors only. The vectors $\mathbf{e}_i$ are chosen in length and direction such that they point exactly onto neighbouring lattice nodes.
Each discrete velocity direction $\mathbf{e}_i$ models a PDF $f_i$, which has its own local equilibrium to relax to. The discretized Boltzmann equation results in
\begin{equation}
    \frac{1}{\Delta t}\left(
            f_i(\mathbf{x}+\mathbf{e}_i\Delta t, t + \Delta t) - f_i(\mathbf{x}, t)
        \right)
        =
        -\frac{1}{\tau}\left(
            f_i(\mathbf{x}, t) - f^\text{eq}_i(\mathbf{x}, t)
        \right).
\end{equation}
The equilibrium distribution $f_i^\text{eq}$ is approximated linearly for an advection diffusion process. For a single grid node, a scalar quantity $\Phi$ with constant background velocity $\mathbf{u}$ on that grid node determines the equilibrium distribution in the $i$-th discrete velocity direction by \cite{mohamad2011lattice}
\begin{equation}\label{eq:feq}
    f_i^\text{eq} = \underbrace{w_i  \left(1 + \frac{\mathbf{u}\cdot \mathbf{e}_i}{c_s^2}\right)}_{\equiv k_i}\Phi.
\end{equation}
Here, the diffusivity $D$ enters the equilibrium distribution via the speed of sound $c_s$ that determines the discretization weights $w_i$. This allows to calculate the change of the PDF $\hat{f}_i$ for the next time step due to \textit{collision}
\begin{equation}\label{eq:fhat}
    \hat{f}_i(\mathbf{x}, t)
    =
    \left( 1  - \frac{\Delta t}{\tau}\right) f_i(\mathbf{x}, t)
    + \frac{\Delta t}{\tau} f^\text{eq}_i(\mathbf{x}, t),
\end{equation}
and the \textit{streaming} to neighbouring nodes for the next time step $t + \Delta t$ for every updated velocity direction $\hat{f}_i$
\begin{equation}
    f_i(\mathbf{x}+\mathbf{e}_i\Delta t, t + \Delta t)
    =
    \hat{f}_i(\mathbf{x}, t).
\end{equation}
With the updated distribution functions on each node, the \textit{macroscopic} scalar quantity on each node can by re-calculated by the sum of the distribution functions $f_i$ \cite{mohamad2011lattice}
\begin{equation}
    \Phi = \sum_i f_i.
\end{equation}

\subsubsection{Diffusion and Lattice Boltzmann weights}
\label{sec:model_D_by_wi}

For the velocity distribution functions in different velocity directions, the equilibrium function of each discrete velocity direction $f^\text{eq}_i$ scales with a weighting factor $w_i$. These weighting factors are obtained by evaluating moment equations of different orders to ensure conservation properties \cite{mohamad2011lattice}. The resulting equations relate the weighting factors with the speed of sound, and therefore indirectly the diffusion constant, since the diffusion is given by \cite{kruger2017lattice}
\begin{equation}
    D = c_s^2 \left(\tau - \frac{\Delta t}{2}\right).
\end{equation}
The diffusion is typically modeled with the relaxation time $\tau$, but for a simulation that is scaled in such a way that the distribution functions relax fully to their equilibrium within one time step ($\tau = \Delta t = 1$), the diffusion can be modeled by scaling the speed of sound $c_s^2$ of the simulation. To scale the speed of sound, the weighting factors $w_i$ are adjusted accordingly. With sets of weighting factors other than the standard set, the moment equations may only be fulfilled up to a certain order. Depending on the physical system, this may or may not violate conservation symmetries, depending on the given order of symmetry that our system shows \cite{mohamad2011lattice}. \\
For our linear, isotropic advection-diffusion problem, we test on the standard weight set and additionally on weight sets that fulfill the moment equations up to order four, which allows us to model different diffusion constants.

\subsection{Quantum state encoding and initial state}
\label{sec:enc_and_init_state}
To encode the flow field with $N$ grid points in $Q$ velocity directions, amplitude encoding is used to generate the quantum state that the QLBM algorithm will operate on.
In order to make the algorithm work for multiple time steps with collision, streaming and calculation of the macroscopic quantities, we build on an initial state vector that contains the scalar grid information only in the first velocity direction $f_1$ while all other velocity directions have zero state amplitudes.
So we explicitly choose an initial state, where all the scalar grid values are stored in the first velocity component, so for a grid node this means $\Phi = f_1$, and the full grid states is
\begin{equation}\label{eq:state_init}
F =
    \begin{pmatrix}
        f_1^1\\
        \vdots\\
        f_1^N\\
        f_2^1\\
        \vdots\\
        f_2^N\\
        f_Q^1\\
        \vdots\\
        f_Q^N
    \end{pmatrix}
    =
    \begin{pmatrix}
        \Phi^1\\
        \vdots\\
        \Phi^N\\
        0\\
        \vdots\\
        0\\
        0\\
        \vdots\\
        0
    \end{pmatrix},
\end{equation}
which will also be referred in short notation grouping the grid into separate \textit{velocity direction subspaces}:
\begin{equation}\label{eq:state_init_short_notation}
    F = (f_1,\ f_2,\ \hdots,\ f_Q)^T = (\Phi,\ 0,\ \hdots,\ 0)^T.
\end{equation}
Note that for $N$ grid points, we need at least $\lceil \log_2(N) \rceil$ qubits, generating $\lceil \log_2(N) \rceil^N$ grid points, where $\lceil \cdot \rceil$ is the ceiling function.
Similarly, for a stencil with $Q$ velocity directions, we need as many direction qubits $\# q_\text{dir}$, such that they generate as many states as is at least the number of velocity directions needed, so
\begin{equation}
    N_Q = 2^{\# q_\text{dir}} \geq Q,
\end{equation}
which will expand the state vector from dimension $N\cdot Q$ to $N \cdot N_Q$, i.e. $(f_1,\ f_2,\ \hdots,\ f_{N_Q})^T = (\Phi,\ 0,\ \hdots,\ 0)^T$, with $N_Q - Q$ zero entries in the short velocity direction subspace notation. \\
In this paper, D1Q2 and D1Q3 as well as D2Q9 stencils are used. They are arranged as shown in the following equations, where each $f_i$ represents a group of $N$ gird points in direction $i$. For D1Q2, the state es encoded by one  velocity qubit to span the two velocity directions
\begin{equation}\label{eq:D1Q2_enc}
    F_\text{D1Q2} = (f_1, f_2)^T,
\end{equation}
for D1Q3, 2 velocity qubits generate the state given by
\begin{equation}\label{eq:D1Q3_enc}
    F_\text{D1Q3} = (f_1, 0, f_2, f_3)^T
\end{equation}
and for D2Q9, 4 velocity qubits (2 for the $x$- and 2 for the $y$-direction) generate the space for the 9  velocity directions arranged by
\begin{equation}\label{eq:D2Q9_enc}
    F_\text{D1Q3} = (f_1, 0, f_2, f_3, 0, 0, 0, 0, f_4, 0, f_6, f_7, f_5, 0, f_8, f_9)^T.
\end{equation}
The arrangements are represented in tables  \ref{tab:D1Q2_arrangement}, \ref{tab:D1Q3_arrangement} and \ref{tab:D2Q9_arrangement}.

\begin{table}[!]
    \centering
    \caption{Arrangement of the D1Q2 velocity direction subspaces with one direction qubit.}
    \begin{tabular}{c|c|c}
        $\ket{q}_{\text{dir},x}$ & $\ket{0}$ & $\ket{1}$ \\ \hline
        direction $f_i$ & $f_1$ & $f_2$ \\
    \end{tabular}
    =
    \begin{tabular}{c|c}
         $\ket{0}$ & $\ket{1}$  \\ \hline
          $\rightarrow$ & $\leftarrow$ \\
    \end{tabular}
    \label{tab:D1Q2_arrangement}
\end{table}

\begin{table}[!]
    \centering
    \caption{Arrangement of the D1Q3 velocity direction subspaces with two direction qubits.}
    \begin{tabular}{c|c|c|c|c}
        $\ket{q_{2}q_{1}}_{\text{dir},x}$ & $\ket{00}$ & $\ket{01}$ & $\ket{10}$ & $\ket{11}$ \\ \hline
        direction $f_i$ & $f_1$ & - & $f_2$ & $f_3$ \\
    \end{tabular}
    =
    \begin{tabular}{c|c|c|c}
         $\ket{00}$ & $\ket{01}$ & $\ket{10}$ & $\ket{11}$ \\ \hline
         rest & - & $\rightarrow$ & $\leftarrow$ \\
    \end{tabular}
    \label{tab:D1Q3_arrangement}
\end{table}

\begin{table}[!]
    \centering
    \caption{Arrangement of the D2Q9 velocity directions for the two $x$-direction qubits denoted by $\ket{q_2 q_1}_x$ and the two $y$-direction qubits denoted by $\ket{q_4 q_3}_y$, respectively.}
    \begin{tabular}{c|c|c|c|c}
         & $\ket{00}_x$ & $\ket{01}_x$ & $\ket{10}_x$ & $\ket{11}_x$ \\ \hline
        $\ket{00}_y$ & $f_1$ & - & $f_2$ & $f_3$ \\ \hline
        $\ket{01}_y$ & - & - & - & - \\ \hline
        $\ket{10}_y$ & $f_4$ & -& $f_6$ & $f_7$ \\ \hline
        $\ket{11}_y$ & $f_5$ & -& $f_8$ & $f_9$ \\
    \end{tabular}
    $=$
    \begin{tabular}{c|c|c|c|c}
        & $\ket{00}_x$ & $\ket{01}_x$ & $\ket{10}_x$ & $\ket{11}_x$ \\ \hline
        $\ket{00}_y$ & rest & - & $\rightarrow$ & $\leftarrow$ \\ \hline
        $\ket{01}_y$ & - & - & - & - \\ \hline
        $\ket{10}_y$ & $\uparrow$  & - & $\nearrow$ & $\nwarrow$ \\ \hline
        $\ket{11}_y$ & $\downarrow$ & - & $\searrow$ & $\swarrow$ \\
    \end{tabular}
    \label{tab:D2Q9_arrangement}
\end{table}

\subsection{The Quantum Lattice Boltzmann Method}

To solve the advection-diffusion equation \eqref{eq:ADE} with a quantum algorithm, the individual steps of the lattice Boltzmann method are solved by quantum algorithms, i.e. collision, streaming and updating the macroscopic values as shown in figure \ref{fig:QLBM_circ_single_step}. \\
The data is encoded in the complex amplitude coefficients of the states of the grid qubits $\ket{q_\text{grid}}$, which for $N$ grid qubits generate space for $2^N$ scalar values. For multiple dimensions, it is separated into groups of grid qubits for each dimension. For the collision operation, additional qubits $\ket{q_\text{dir}}$ are added, to enable to duplicate the grid in order to operate for the different LBM directions, given by the specific choice of the velocity set (LBM stencil). Also the streaming is then performed conditioned on the different direction qubits to transport in the corresponding directions. The macroscopic step finally merges the different directions to recalculate the new macroscopic quantities to update for a new local scalar quantity and local equilibrium. This QLBM routine generally requires all velocity subspaces but the first one to have zero probability amplitude in order to obtain the correct assignment by the collision step, as will be discussed in further subsections. Hence, the \textit{reinitialization QLBM} algorithms reinitialize the state vector accordingly in each time step. \\
Our goal is to construct a scheme for a QLBM step that does not require a state extraction and reinitialization in between the time steps. Therefore, we propose a new extended scheme with a {\sc re-prep} block and additional time qubits $\ket{q_\text{time}}$ as shown in figure \ref{fig:QLBM_circ_mult_steps}, to enable a fully quantum algorithm for all time steps from start to end without a measurement of the state at any time in between.

\begin{figure}[!htbp]
    \centering

    \begin{quantikz}[wire types={b,b},classical gap=0.07cm]
        \lstick{$\ket{q_\text{grid}}$} &\gate[2]{\text{\sc collision}} & \gate[2]{\rightarrow}\gategroup[2,steps=4,style={dashed,rounded corners, inner xsep=2pt},background,label style={label
position=below,anchor=north,yshift=-0.2cm}]{\sc{streaming}}& \gate[2]{\leftarrow} &  \ \ldots \ & \gate[2]{...} & & \\
        \lstick{$\ket{q_\text{dir}}$} & & & &  \ \ldots \ & & \gate{\text{\sc macros}} &
    \end{quantikz}

    \caption{Quantum circuit for a single quantum lattice Boltzmann time step.}
    \label{fig:QLBM_circ_single_step}
\end{figure}
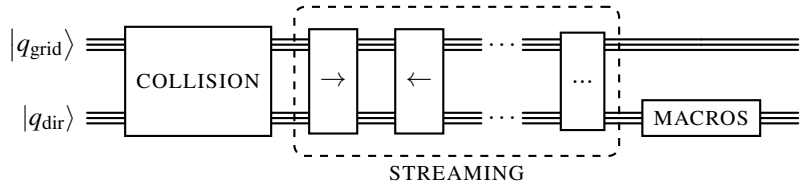

\begin{figure}[!htbp]
    \centering


    \begin{quantikz}[wire types={b,b,b},classical gap=0.1cm]
        \lstick{$\ket{q_\text{grid}}$} &\gate[2]{\text{\sc collision}} & \gate[2]{\rightarrow}\gategroup[2,steps=4,style={dashed,rounded corners, inner xsep=2pt},background,label style={label
position=below,anchor=north,yshift=-0.2cm}]{\sc{streaming}}& \gate[2]{\leftarrow} &  \ \ldots \ & \gate[2]{...} & & & \\
        \lstick{$\ket{q_\text{dir}}$} & & &  &  \ \ldots \ &  & \gate{\text{\sc macros}} & \gate[2]{\text{\sc re-prep}} & \\[0.4cm]
        \lstick{$\ket{q_\text{time}}$} & & & & & & & &
    \end{quantikz}

    \caption{Our quantum circuit extension of the QLBM routine for multiple quantum lattice Boltzmann time steps without state extraction or reinitialization.}

    \label{fig:QLBM_circ_mult_steps}
\end{figure}
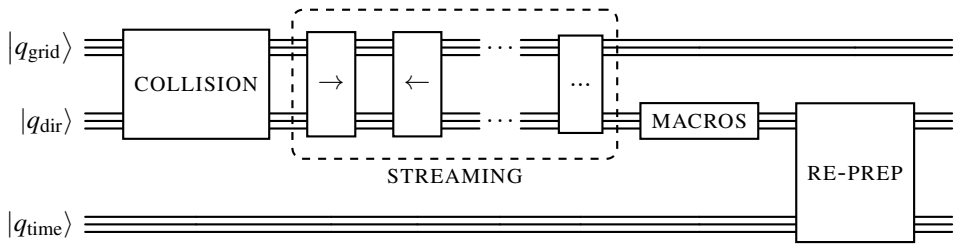

\begin{figure}[!htbp]
    \centering

    \begin{quantikz}[wire types={b,b,b,q},classical gap=0.1cm, column sep=0.4cm, row sep=0.8cm]
        \lstick{$\ket{q_\text{grid}}$} &
            \gate[3,disable auto height]{\rotatebox{90}{\text{\sc Initialize}}} & \gate[2,disable auto height]{\rotatebox{90}{\text{\sc collision}}}
            \gategroup[3,steps=4,style={solid,rounded corners, inner xsep=2pt},background,label style={label position=below,anchor=north,yshift=-0.2cm}]{\text{repeat $T$ times
            }}
             & \gate[2,disable auto height]{\rotatebox{90}{\text{\sc streaming}}} &
            & & & & \meter{} \\
        \lstick{$\ket{q_\text{dir}}$} & & &  & \gate[1]{\rotatebox{90}{\text{\sc macros}}} & \gate[2,disable auto height]{\rotatebox{90}{\text{\sc re-prep}}} & & & \meter{} \\
        \lstick{$\ket{q_\text{time}}$} & & & & & & & & \meter{} 
    \end{quantikz}

    \caption{Full quantum circuit of the QLBM simulation with the extension of the QLBM routine for $T$ time steps without state extraction or reinitialization from figure \ref{fig:QLBM_circ_mult_steps}.}
    \label{fig:QLBM_full_circ}
\end{figure}
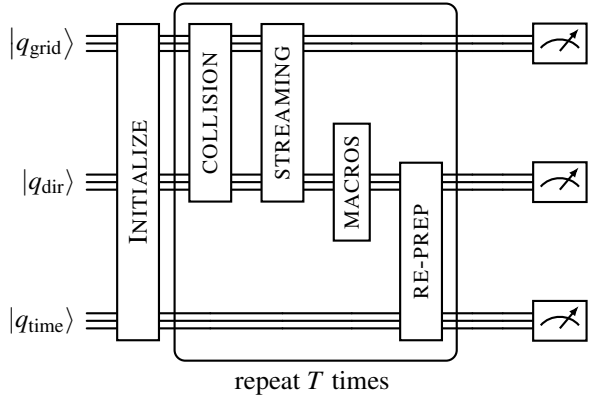

\subsubsection{The collision step}
\label{sec:collision}

The collision step in the context of the LBM determines how each of the velocity probability distribution function $f_i$ relaxes towards its local equilibrium for that specific velocity direction $i$. Assuming a full relaxation within a time step, i.e $\Delta t / \tau = 1$, the collision step is directly determined by the equilibrium distribution $f^\text{eq}$ in equation \eqref{eq:feq}. Although the diffusion usually dictates the relaxation time $\tau$, this does not restrict the diffusion to a fixed value because we can model the diffusion by the modeled speed of sound $c_s$ and weights $w_i$ as described in section \ref{sec:model_D_by_wi}. \\
For $N$ grid points and $Q$ velocity directions, our total state has dimension $N\cdot Q$, for which we have to calculate the relaxation for each entry. Collecting this in a state vector $F = (f_1^1,\ \dots,\ f_1^N,\ \dots,\ f_Q^1,\ \dots,\ f_Q^N)^T$, the collision operation for each $f_i^n$, $i \in [1, Q]$, $n \in [1, N]$ as dictated by equation \eqref{eq:fhat} can be written as the linear system
\begin{equation}
    \hat{F} = AF,
\end{equation}
where $A$ is the diagonal collision matrix. The collision matrix is generally not unitary and thus can not be decomposed into a series of available unitary operations of a quantum computer. Therefore, Budinski \cite{budinski2021quantum} separates $A$ into a sum of diagonal unitary matrices $A = (B_1 + B_2)/2 $ with $B_{1,k,k} = A_{k,k} + i\sqrt{1 - A_{k,k}^2}$ and $B_{2,k,k} = A_{k,k} - i\sqrt{1 - A_{k,k}^2}$, following a linear combination of unitaries (LCU) approach. The idea of the LCU approach is to add an additional ancilla qubits \ket{a} and duplicate the grid coefficients up to a norm factor into the new ancilla generated subspace. Now, $B_1$ and $B_2$ can operate separately on the two subspaces by conditioning the operations on the ancilla state \ket{a}. To do so, $B_1$ operates conditioned on $\ket{a} = \ket{0}$ and while $B_2$ is performed conditioned where $\ket{a} = \ket{1}$ and the results in the two subspaces are summed in a last step (for summation, cf. the summation in section \ref{sec:macro}). \\
After the LCU summation step, there remains a significant amount of amplitude of the complex amplitude coefficient of the states in the additional LCU ancilla subspace, that results in a loss of probability that we want to avoid for our algorithm. So instead of using the LCU, we want to mimic the distribution into the multiple directions dictated by $A$ by a unitary collision operation $U_\text{coll}$ in the first place.\\
\newline
The idea is to start a time step in a state that has all the scalar information in one velocity direction subspace (cf. section \ref{sec:enc_and_init_state}) and then mimic the distribution of the collision operation along the velocity directions by a unitary matrix. To achieve starting in this state, the state is re-prepared at the end of each time step by the {\sc re-prep} block as described in section \ref{sec:re-prep} and as sketched in figure \ref{fig:QLBM_circ_mult_steps}. The choice of this state is made possible by our choice of $\Delta t / \tau = 1$ because after the collision the state relaxes fully into the equilibrium state, which only depends on $\Phi$ of the node, independent of how $\Phi$ is distributed along the different directions in the first place (recap $\Phi = \sum_i f_i$ for a specific grid point). So if $\Phi$ is fully captured by one velocity direction, we can make sure that our desired operation $U_\text{coll}$ operates on these states according to the corresponding entries in $A$, while the remaining entries can be chosen arbitrarily, but in particular such that our desired matrix $U_\text{coll}$ becomes unitary. \\
\newline
To find a general collision description, we start with the simplest QLBM stencil and generalize to arbitrary stencils further in the subsection. The simplest case for a QLBM collision is given by a \textbf{D1Q2} stencil, so by one spacial dimension and two velocity directions $Q=2$, leading to a state vector $F = (f_1^1,\ \dots,\ f_1^N,\ f_2^1,\ \dots,\ f_2^N)^T$. Here, the collision matrix $A$ has the following form:
\begin{equation}
A =
\begin{pNiceArray}{cc|cc}
  \Block{2-2}{{A_1}} & \   & \Block{2-2}{{A_2}} \\
  \ & \ \\
  \hline
  \Block{2-2}{{A_3}} && \Block{2-2}{{A_4}} \\
  && \  & \
\end{pNiceArray},
\end{equation}
where the $A_i$ are diagonal $N \times N$ matrices. With the state prepared as in equation \eqref{eq:state_init}, $A_2$ and $A_4$ do not contribute to the collision calculation. Additionally, according to equation \eqref{eq:feq}, all the diagonal entries $a_i$ inside an $A_i$ are all equal and we can write:
\begin{equation}
    A =
    \begin{pNiceArray}{ccc|ccc}
        a_1 & & & a_2 & & \\
        & \ddots & & & \ddots &\\
        & & a_1 & & & a_2\\
    \hline
        a_3 & & & a_4 & & \\
        & \ddots & & & \ddots &\\
        & & a_3 & & & a_4\\
    \end{pNiceArray}
    =
    \begin{pmatrix}
       a_1 & a_2 \\
       a_3 & a_4 \\
    \end{pmatrix}
    \otimes
    \mathbb{1}_N.
\end{equation}
This can be represented by a quantum operation using a $2 \times 2 $ rotational gate, the $RY(\theta)$-gate, on one qubit and no operation ($\equiv$ identity operation $\mathbb{1}$) on $N$ qubits. So the rotation operation has to fit the coefficients $a_i$, but only for $a_1$ and $a_3$, since our state is prepared in such a way that $a_2$ and $a_4$ do not contribute
\begin{equation}
    \begin{pNiceArray}{>{\strut}cc}[margin,extra-margin = 1pt]
    a_1 & a_2 \\
    a_3 & a_4
    \CodeAfter
      \begin{tikzpicture}
      \node [dashed, draw=black, rounded corners=2pt, inner ysep = 0pt,
           rotate fit=0, fit = (1-1) (2-1) ] {} ;
      \end{tikzpicture}
    \end{pNiceArray}
    =
    \begin{pNiceArray}{>{\strut}rr}[margin,extra-margin = 1pt] 
    \cos(\theta / 2) & -\sin(\theta / 2) \\
    \sin(\theta / 2) & \cos(\theta / 2)
    \CodeAfter
      \begin{tikzpicture}
      \node [dashed, draw=black, rounded corners=2pt, inner ysep = 0pt,
           rotate fit=0, fit = (1-1) (2-1) ] {} ;
      \end{tikzpicture}
    \end{pNiceArray}
    \equiv RY(\theta),
\end{equation}
leading to the conditions for choosing $\theta$ such that the $a_i$ correspond to map the vector $\Phi$ from equation \eqref{eq:state_init} to $f^\text{eq}$ in equation \eqref{eq:feq} of the corresponding velocity direction:
\begin{align}
    a_1 &= k_1 \equiv w_1 \left(1 + \frac{u\cdot e_1}{c_s^2}\right) \overset{!}{=} \cos(\theta / 2), \\
    a_3 &= k_2 \equiv w_2 \left(1 + \frac{u\cdot e_2}{c_s^2}\right) \overset{!}{=} \sin(\theta / 2).
\end{align}
So applying $RY(\theta)$ to the state as prepared in equation \eqref{eq:state_init}, this operation effectively shifts a certain fraction from all $\Phi$ in the first velocity direction
to the second velocity direction, such that the ratio of $k_1/(k_1+k_2)$ is kept in the first velocity subspace while moving $k_2/(k_1+k_2)$ into the second subspace. So essentially, we want to choose $\theta$, such that we distribute our state $(\Phi, 0)^T$ to $(k_1 \Phi, k_2 \Phi)^T$, but this can not be archived by unitary operation in general. So instead, we choose $\theta$, such that we keep the proportion of $k_1/k_2$ for $(\cos(\theta / 2) \Phi, \sin(\theta / 2) \Phi)^T$.
This leads to our condition and single solution for the argument $\theta$:
\begin{equation}\label{eq:theta_D1Q2}
    \frac{k_1}{k_2} = \frac{\cos(\theta / 2)}{\sin(\theta / 2)}
    \Longleftrightarrow
    \theta = 2 \arccos \underbrace{\left( \frac{k_1}{\sqrt{ k_1^2 + k_2^2}} \right)}_{\equiv n}
    = 2 \arcsin \left( \frac{k_2}{\sqrt{ k_1^2 + k_2^2}} \right).
\end{equation}
Note, that instead of calculating the ratio of $k_1$ and $k_2$, only a normalized ratio with $n \equiv k_1/\sqrt{k_1^2 + k_2^2}$ and $k_2/\sqrt{k_1^2 + k_2^2}$ is calculated, which will lead to a certain decay discussed in section \ref{sec:decay}.
\subsubsection*{Generalization}
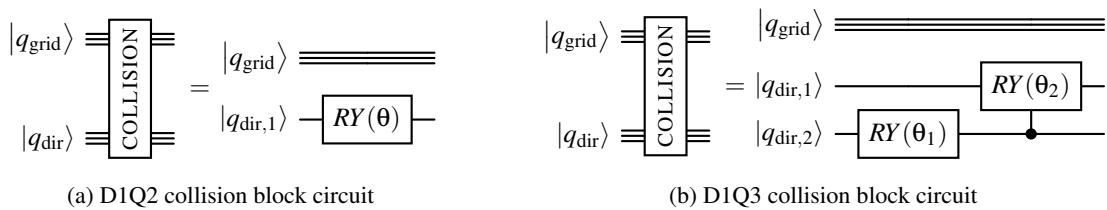
\begin{figure}[!htbp]
    \centering
    \begin{subfigure}{0.49\textwidth}
        \centering
         \begin{quantikz}[wire types={b,b}, classical gap=0.07cm, row sep=0.8cm, column sep=0.3cm]
            \lstick{$\ket{q_\text{grid}}$} & \gate[2,disable auto height]{\rotatebox{90}{\text{\sc collision}}} & \\
            \lstick{$\ket{q_{\text{dir}}}$} & &
        \end{quantikz}
        $=$\begin{quantikz}[wire types={b,q}, classical gap=0.07cm, row sep=0.5cm, column sep=0.3cm]
            \lstick{$\ket{q_\text{grid}}$} & &  \\
            \lstick{$\ket{q_{\text{dir},1}}$} & \gate{RY(\theta)} &
        \end{quantikz}
        \caption{D1Q2 collision block circuit}
        \label{fig:D1Q2_coll_circ}
    \end{subfigure}
    \begin{subfigure}{0.49\textwidth}
        \centering
        \begin{quantikz}[wire types={b,b}, classical gap=0.07cm, row sep=0.8cm, column sep=0.3cm]
            \lstick{$\ket{q_\text{grid}}$} & \gate[2,disable auto height]{\rotatebox{90}{\text{\sc collision}}} & \\
            \lstick{$\ket{q_{\text{dir}}}$} & &
        \end{quantikz}
        $=$\begin{quantikz}[wire types={b,q,q}, classical gap=0.07cm, row sep=0.5cm, column sep=0.3cm]
            \lstick{$\ket{q_\text{grid}}$} & & & \\ 
            \lstick{$\ket{q_{\text{dir},1}}$} &  &  \gate{RY(\theta_2)}  & \\ [-0.5cm]
            \lstick{$\ket{q_{\text{dir},2}}$} & \gate{RY(\theta_1)} &  \ctrl{-1} &
        \end{quantikz}
        \caption{D1Q3 collision block circuit}
        \label{fig:D1Q3_coll_circ}
    \end{subfigure}%

    \caption{Quantum circuit for the collision step in figure \ref{fig:QLBM_circ_mult_steps} for the D1Q2 scheme (fig. \ref{fig:D1Q2_coll_circ}) and a the D1Q3 scheme (fig. \ref{fig:D1Q3_coll_circ}) for spatially constant diffusion and flow velocity. For spatially varying diffusion or flow velocities, multiple $RY$ gates with controls on the grid qubits $\ket{q_\text{grid}}$ need to be used.}
\end{figure}

\begin{figure}[!htbp]
    \centering
     \begin{quantikz}[wire types={b,b}, classical gap=0.07cm, row sep=0.8cm, column sep=0.3cm]
        \lstick{$\ket{q_\text{grid}}$} & \gate[2,disable auto height]{\rotatebox{90}{\text{\sc collision}}} & \\
        \lstick{$\ket{q_{\text{dir}}}$} & &
    \end{quantikz}
    $=$\begin{quantikz}[wire types={b,q,q,q,q}, classical gap=0.07cm, row sep=-0.1cm, column sep=0.1cm]
            \lstick{$\ket{q_\text{grid}}$} \hspace{0.4cm} & & & & & & & & & \\ [0.5cm]
            \lstick{$\ket{q_{\text{dir},1}}$} \hspace{0.4cm} & &
            \gate{\underset{{\ket{2}\rightarrow\ket{3}}}{RY}} & \octrl{1} & & \octrl{1} & \ctrl{1} & \octrl{1} & \ctrl{1} & \\
            \lstick{$\ket{q_{\text{dir},2}}$} \hspace{0.4cm} &  \gate{\underset{{\ket{0}\rightarrow\ket{2}}}{RY}} & \ctrl{-1} & \octrl{2} & & \ctrl{2} & \ctrl{2} & \ctrl{1} & \ctrl{1} & \\
            \lstick{$\ket{q_{\text{dir},3}}$} \hspace{0.4cm} & & & & \gate{\underset{{\ket{8}\rightarrow\ket{12}}}{RY}} & & & \gate{\underset{{\ket{10}\rightarrow\ket{14}}}{RY}} & \gate{\underset{{\ket{11}\rightarrow\ket{15}}}{RY}} & \\
            \lstick{$\ket{q_{\text{dir},4}}$} \hspace{0.4cm} & & & \gate{\underset{{\ket{0}\rightarrow\ket{8}}}{RY}} & \ctrl{-1} & \gate{\underset{{\ket{2}\rightarrow\ket{10}}}{RY}} & \gate{\underset{{\ket{3}\rightarrow\ket{11}}}{RY}} & \ctrl{-1} & \ctrl{-1} &
    \end{quantikz}

    \caption{Quantum circuit for the collision step in figure \ref{fig:QLBM_circ_mult_steps} for the D2Q9 scheme for spatially constant diffusion and flow velocity. Each $RY$-gate distributes the velocity distribution functions into their corresponding velocity direction subspace according to the procedure in table \ref{tab:D2Q9_distributing_f} in section \ref{sec:appendix_coll_distr_D2Q9} the appendix. The argument $\theta$ for each $RY$-gate has to be chosen according to equation \eqref{eq:theta_coll_keep_mult_subspaces}. For additionally spatially varying diffusion or flow velocities, multiple $RY$ gates with controls on the grid qubits $\ket{q_\text{grid}}$ need to be used.}
    \label{fig:D2Q9_coll_circ}
\end{figure}
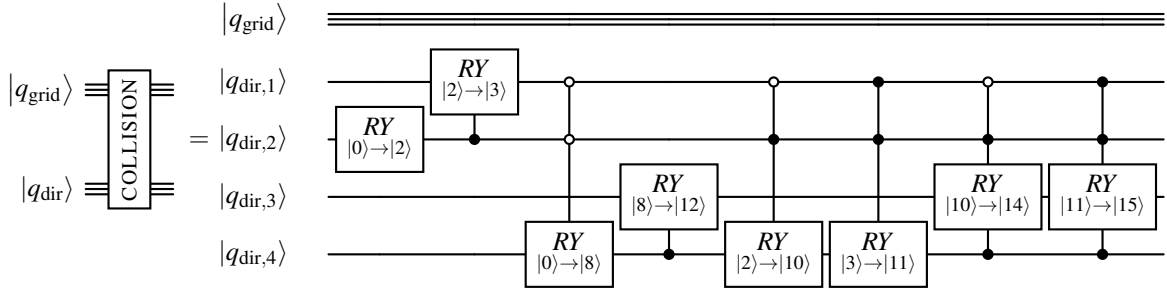

Extending this idea to \textbf{D1Q3} scheme, three velocity directions are modeled ($Q=3$) using $N_Q = 2$ direction qubits that span four velocity direction subspaces. Here, the first velocity direction is the resting node direction $(f_1^{1},\  \hdots, f_1^{N})$, while the second and third direction are the left and right streaming direction, respectively, which are arranged as shown in table \ref{tab:D1Q3_arrangement}. In order to include the additional direction, the same idea is used, starting with an initial state vector as in equation \eqref{eq:state_init} and sequentially distributing to the additional direction subspaces.
Now two steps are modeled: in the first step, an angle $\theta_1$ is determined, such that the amount $n_1$ is kept in the first (resting) direction while all remaining fraction $n'_1$ is shifted to the second direction subspace $(\Phi, 0, 0, 0)^T \rightarrow (n_1 \Phi, 0, n'_1 \Phi, 0)^T$. Note that $n_1$ is the normalized correct amount of the first subspace while the second subspace now contains all remaining information, so for the second and third velocity direction. This remaining information still needs to be further distributed. Now by a second $RY$ operation, an angle $\theta_2$ is determined to distribute between the second and third subspace, according to the LBM weights: $(n_1 \Phi, 0, n'_1 \Phi, 0)^T \rightarrow (n_1 \Phi, 0, n_2 \Phi, n_3 \Phi)^T$. This results in two rotation operations with angles
\begin{align}
    \theta_{1} = 2 \arccos \underbrace{\left( \frac{k_1}{\sqrt{ k_1^2 + k_2^2 + k_3^2}} \right)}_{\equiv n_{1}} , \ \ \
    \theta_{2} = 2 \arccos \underbrace{\left( \frac{k_2}{\sqrt{ k_2^2 + k_3^2}} \right)}_{\equiv n_{2}}.
\end{align}
For a 1D stencil, we shift all the information sequentially through each of the subspaces while keeping a fraction according the LBM weights in the corresponding subspace \linebreak[4] $(\Phi, 0, \hdots, 0)^T \rightarrow (n_1 \Phi, n' \Phi, \hdots, 0)^T \rightarrow \hdots \rightarrow (n_1 \Phi, n_2 \Phi, \hdots, n_Q \Phi)^T$, where each operation is performed with a general angle for each of the rotation gates of
\begin{equation}\label{eq:theta_coll_DxQx}
    \theta_{i} = 2 \arccos \underbrace{\left( \frac{k_i}{\sqrt{ \sum_{j=i}^Q k_j^2 }} \right)}_{\equiv n_{i}} .
\end{equation}
For multiple spacial dimensions, the sequential distribution of the signal through the velocity subspaces may have to be adjusted.
If the signal may have to be kept for multiple directions $k_j$ in a certain subspace (as will be necessary in our D2Q9 distribution procedure), the value of $k_i$ has to be replaced by the Euclidean norm of all the directions that we want to keep in the subspace:
\begin{equation}\label{eq:theta_coll_keep_mult_subspaces}
    \theta_{i} = 2 \arccos \left( \frac{
        \sqrt{ \sum_{\text{keep}} k_\text{keep}^2 }
    }{
        \sqrt{ \sum_{\text{inv}} k_\text{inv}^2 }
    } \right),
\end{equation}
where $k_\text{keep}$ are the $k_i$ of the directions that (temporarily) remain in the direction subspace and $k_\text{inv}$ are all the subspaces involved in the operation, so those which are shifted to another subspace and those which will remain in the respective subspace.\\
\newline
In this work, the D1Q2, D1Q3 and D2Q9 stencils are used. The details of the quantum circuit of the collision block from figure \ref{fig:QLBM_circ_mult_steps} is shown in figure \ref{fig:D1Q2_coll_circ} and \ref{fig:D1Q3_coll_circ} for the D1Q2 and the D1Q3 scheme, respectively, and for the D2Q9 scheme it is shown in figure \ref{fig:D2Q9_coll_circ}. A spatially constant diffusion and flow velocity is assumed, so there is no dependence on the grid qubits. For a spatially varying diffusion, additional $RY$ gates with controls on the grid qubits need to be used. \\
To represent the two velocity directions in the D1Q2 scheme, we use one direction qubit in the $\ket{q_\text{dir}}$ register to span the state vector in equation \eqref{eq:state_init} of length $2N$, where we already use $\log_2(N)$ qubits for the representation of the $N$ grid points. \\
For the D1Q3 scheme, we need a state vector in equation \eqref{eq:state_init} of length $3N$, so we need an second qubit in the direction register, as explained in section \ref{sec:enc_and_init_state}. This will span our state space to $4 N$.\\
For the D2Q9 scheme, it is not feasible to distribute the velocity directions in a sequential way, always shifting all the signal to the next direction subspace.
Specifically, for the D2Q9 scheme used in this work, first the amplitude from the resting subspace will be distributed along the $x$-direction subspaces as for D1Q3, but in a second series of steps, the direction subspaces superposing the $x$-direction (including the resting direction) with up and down directions will get their corresponding signal. The details of the order on how the directions are distributed to their corresponding subspace is shown in the appendix in section \ref{sec:appendix_coll_distr_D2Q9} in table \ref{tab:D2Q9_distributing_f}. For the distribution procedure, a set of eight $RY(\theta)$-gate operations are used and $N_Q = 4$ direction qubits generating 16 subspaces are used to accommodate for the $Q = 9$ velocity directions. To perform the operations only on the desired subspace, several controls are set to the $RY$-gates, which are shown in the collision circuit in figure \ref{fig:D2Q9_coll_circ}.

\newpage
\subsubsection{The streaming step}

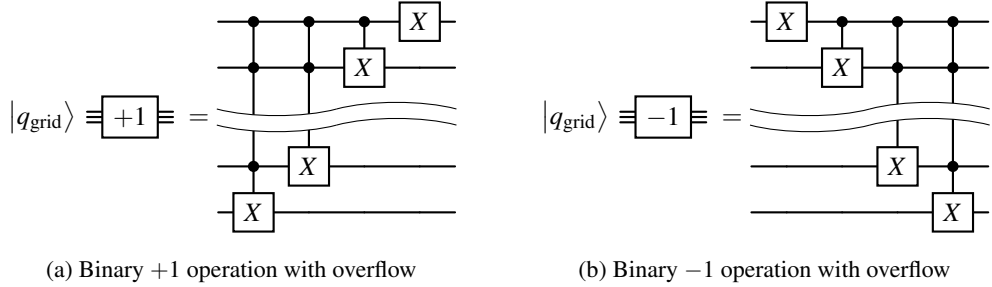
\begin{figure}[!htbp]
    \centering
    \begin{subfigure}{0.49\textwidth}
        \centering
        \begin{quantikz}[wire types={b}, classical gap=0.07cm, column sep=0.2cm]
           \lstick{$\ket{q_\text{grid}}$} & \gate{+1} &
        \end{quantikz}
        $=$\begin{quantikz}[column sep=0.2cm, row sep=0.1cm]
           & \ctrl{4} & \ctrl{3} & \ctrl{1} & \gate{X} & \\
           & \control{} & \control{} & \gate{X} &  & \\ [0.3cm]
           \wave&&&&&  \\ [0.3cm]
           & \control{} & \gate{X} &  &  & \\
           & \gate{X} & & &  &
        \end{quantikz}
        \caption{Binary $+1$ operation with overflow}
        \label{fig:+1_circ}
    \end{subfigure}
    \hspace*{-1cm}   
    \begin{subfigure}{0.49\textwidth}
        \centering
        \begin{quantikz}[wire types={b}, classical gap=0.07cm, column sep=0.2cm]
        \lstick{$\ket{q_\text{grid}}$} & \gate{-1} &
        \end{quantikz}
        $=$\begin{quantikz}[column sep=0.2cm, row sep=0.1cm]
           & \gate{X} & \ctrl{1} & \ctrl{3} & \ctrl{4} & \\
           &  & \gate{X} &  \control{} & \control{} & \\ [0.3cm]
           \wave&&&&&  \\ [0.3cm]
           &  &  & \gate{X} & \control{} & \\
           &  & & & \gate{X} &
        \end{quantikz}
        \caption{Binary $-1$ operation with overflow}
        \label{fig:-1_circ}
    \end{subfigure}%

    \caption{Binary $+1$ and $-1$ operations with overflow as the basis of the right, left, up and down operations in the QLBM circuit in figure \ref{fig:QLBM_circ_mult_steps}.}
    \label{fig:+1_and_-1_operations}
\end{figure}

The streaming for the stencils used in this work as indicated in the circuit in figure \ref{fig:QLBM_circ_mult_steps} are basic periodic right ($\rightarrow$) and left ($\leftarrow$) shift operations in 1D and additional up ($\uparrow$) and down ($\downarrow$) operations including diagonal superpositions of them, respectively, in 2D. This is done by a binary $+1$ and $-1$ operation with binary integer overflow, conditioned on the direction qubits. The quantum circuit for the binary shift operations is adapted from Todorova et al. \cite{todorova2020quantum} as presented in figures \ref{fig:+1_circ} and \ref{fig:-1_circ}, where instead of changing the condition state of the control qubits for $-1$ compared to $-1$, we instead reverse the order of the gates of $+1$ to represent the $-1$ operation.

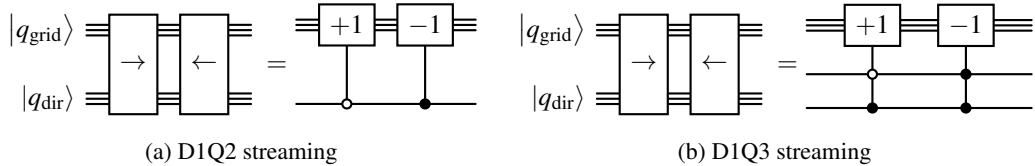
\begin{figure}[!htbp]
    \centering
    \begin{subfigure}{0.49\textwidth}
        \centering
        \begin{quantikz}[wire types={b,b}, classical gap=0.07cm, column sep=0.3cm]
           \lstick{$\ket{q_\text{grid}}$} & \gate[2]{\rightarrow} & \gate[2]{\leftarrow} &\\
           \lstick{$\ket{q_\text{dir}}$} & & &
        \end{quantikz}
        $=$\begin{quantikz}[wire types={b,q}, classical gap=0.07cm, row sep=0.7cm, column sep=0.3cm]
            & \gate{+1} & \gate{-1} & \\
            & \octrl{-1} &\ctrl{-1} &
        \end{quantikz}
        \caption{D1Q2 streaming}
        \label{fig:D1Q2_streaming_circ}
    \end{subfigure}
    \hspace*{-1cm}   
    \begin{subfigure}{0.49\textwidth}
        \centering
        \begin{quantikz}[wire types={b,b}, classical gap=0.07cm, column sep=0.3cm]
           \lstick{$\ket{q_\text{grid}}$} & \gate[2]{\rightarrow} & \gate[2]{\leftarrow} &\\
           \lstick{$\ket{q_\text{dir}}$} & & &
        \end{quantikz}
        $=$\begin{quantikz}[wire types={b,q,q}, classical gap=0.07cm, align equals at = 1.8]
            & \gate{+1} & \gate{-1} & \\ [-0.2cm]
            & \octrl{-1} &\ctrl{-1} & \\ [-0.2cm]
            & \ctrl{-1} &\ctrl{-1} &
        \end{quantikz}
        \caption{D1Q3 streaming}
        \label{fig:D1Q3_streaming_circ}
    \end{subfigure}%

    \caption{Arrangement of the controls on the direction qubits for the streaming operations of the D1Q2 scheme \eqref{fig:D1Q2_streaming_circ} and D1Q3 scheme \eqref{fig:D1Q3_streaming_circ}.}
    \label{fig:L_and_R_operations}
\end{figure}

To perform the correct streaming direction in the corresponding direction subspace, the $+1$ and $-1$ streaming operations are conditioned on the direction qubits, which depends on the velocity set chosen by the QLBM stencil. The conditions for our choice of direction subspaces as described in section \ref{sec:enc_and_init_state} are shown in figure \ref{fig:L_and_R_operations}, so every gate of the $+1$ and $-1$ operation in figure \ref{fig:+1_and_-1_operations} has to have additional controls on the direction qubits. The controls are chose in such a way that they shift the correct subspace of the state, which for the D1Q2 scheme is chosen to $(f_1, f_2)^T = (f_{\rightarrow}, f_{\leftarrow})^T$ and for the D1Q3 is chosen to be $(f_1, 0, f_2, f_3)^T = (f_\text{rest}, 0, f_{\rightarrow}, f_{\leftarrow})^T$ (cf. tables \ref{tab:D1Q2_arrangement} and \ref{tab:D1Q3_arrangement}).

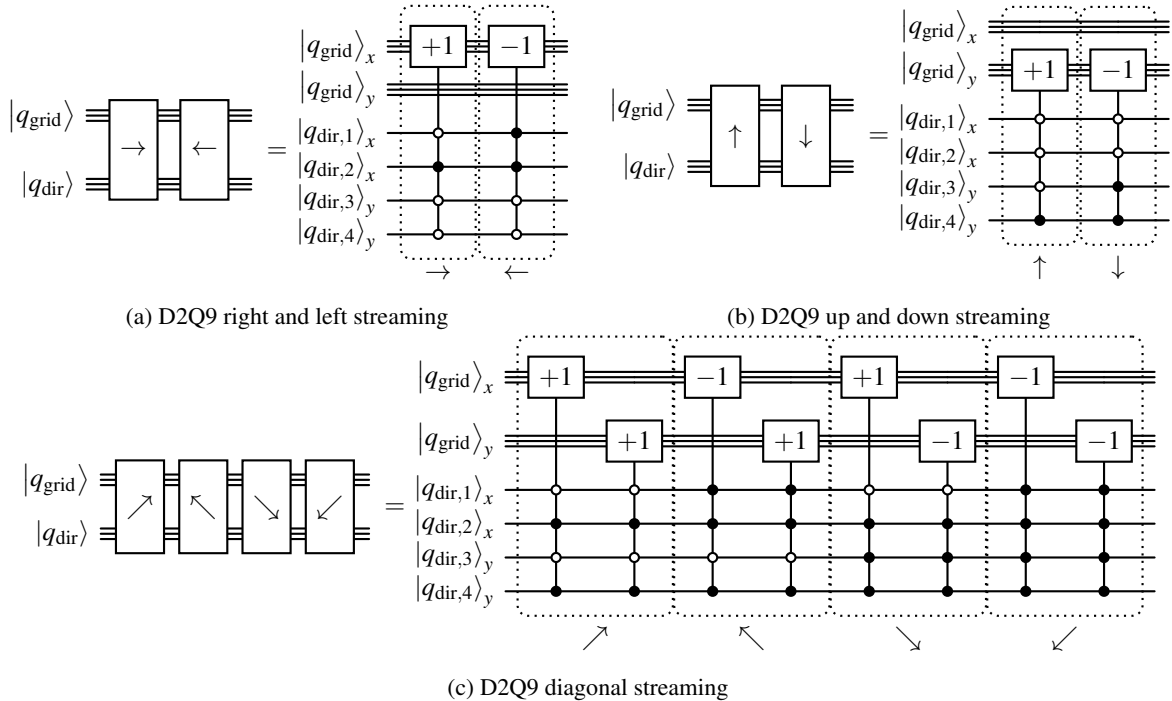
\begin{figure}[!htbp]
    \centering
    \begin{subfigure}{0.49\textwidth}
        \centering
        \begin{quantikz}[wire types={b,b}, classical gap=0.07cm, column sep=0.3cm]
           \lstick{$\ket{q_\text{grid}}$} & \gate[2]{\rightarrow} & \gate[2]{\leftarrow} &\\
           \lstick{$\ket{q_\text{dir}}$} & & &
        \end{quantikz}
        $=$\begin{quantikz}[wire types={b,b,q,q,q,q}, classical gap=0.07cm,
        row sep=0.3cm, column sep=0.3cm]
           \lstick{$\ket{q_{\text{grid}}}_x$} & \gate{+1}
           \gategroup[6,steps=1,style={dotted,rounded corners, inner xsep=0.1pt},background,label style={label position=below,anchor=north,yshift=-0.2cm}]{\text{$\rightarrow$}} &
           \gate{-1}
           \gategroup[6,steps=1,style={dotted,rounded corners, inner xsep=0.1pt},background,label style={label position=below,anchor=north,yshift=-0.2cm}]{\text{$\leftarrow$}} &
           \\
           \lstick{$\ket{q_{\text{grid}}}_y$} & & & \\ [0.2cm]
           \lstick{$\ket{q_{\text{dir},1}}_x$} & \octrl{-2} & \ctrl{-2} & \\
           \lstick{$\ket{q_{\text{dir},2}}_x$} & \ctrl{-1} & \ctrl{-1} & \\
           \lstick{$\ket{q_{\text{dir},3}}_y$} & \octrl{-1} & \octrl{-1} & \\
           \lstick{$\ket{q_{\text{dir},4}}_y$} & \octrl{-1} & \octrl{-1} &
        \end{quantikz}
        \caption{D2Q9 right and left streaming}
        \label{fig:D2Q9_right_left_streaming_circ}
    \end{subfigure}
    \hspace*{-0.1cm}   
    \begin{subfigure}{0.49\textwidth}
        \centering
        \begin{quantikz}[wire types={b,b}, classical gap=0.07cm, column sep=0.3cm,
        row sep=0.3cm]
           \lstick{$\ket{q_\text{grid}}$} & \gate[2,disable auto height]{\ \uparrow \ } & \gate[2,disable auto height]{\ \downarrow \ } &\\
           \lstick{$\ket{q_\text{dir}}$} & & &
        \end{quantikz}
        $=$\begin{quantikz}[wire types={b,b,q,q,q,q}, classical gap=0.07cm,
        row sep=0.3cm, column sep=0.3cm]
           \lstick{$\ket{q_{\text{grid}}}_x$} &
           \gategroup[6,steps=1,style={dotted,rounded corners, inner xsep=0.1pt},background,label style={label position=below,anchor=north,yshift=-0.2cm}]{\text{$\uparrow$}} &
           \gategroup[6,steps=1,style={dotted,rounded corners, inner xsep=0.1pt},background,label style={label position=below,anchor=north,yshift=-0.2cm}]{\text{$\downarrow$}} &
           \\
           \lstick{$\ket{q_{\text{grid}}}_y$} & \gate{+1} & \gate{-1} & \\
           \lstick{$\ket{q_{\text{dir},1}}_x$} & \octrl{-1} & \octrl{-1} & \\
           \lstick{$\ket{q_{\text{dir},2}}_x$} & \octrl{-1} & \octrl{-1} & \\
           \lstick{$\ket{q_{\text{dir},3}}_y$} & \octrl{-1} & \ctrl{-1} & \\
           \lstick{$\ket{q_{\text{dir},4}}_y$} & \ctrl{-1} & \ctrl{-1} &
        \end{quantikz}
        \caption{D2Q9 up and down streaming}
        \label{fig:D2Q9_up_down_streaming_circ}
    \end{subfigure}
    %
    %
    %
    %
    \begin{subfigure}{0.99\textwidth}
        \centering
        \begin{quantikz}[wire types={b,b}, classical gap=0.07cm, column sep=0.2cm,
        row sep=0.2cm]
            \lstick{$\ket{q_\text{grid}}$} &
            \gate[2,disable auto height]{\text{$\nearrow$}} &
            \gate[2,disable auto height]{\text{$\nwarrow$}} &
            \gate[2,disable auto height]{\text{$\searrow$}} &
            \gate[2,disable auto height]{\text{$\swarrow$}} &\\
            \lstick{$\ket{q_\text{dir}}$} & & & & &
        \end{quantikz}
        $=$\begin{quantikz}[wire types={b,b,q,q,q,q}, classical gap=0.07cm,
        row sep=0.3cm, column sep=0.3cm]
            \lstick{$\ket{q_{\text{grid}}}_x$} &
            \gate{+1}
            \gategroup[6,steps=2,style={dotted,rounded corners, inner xsep=0.5pt},background,label style={label position=below,anchor=north,yshift=-0.2cm}]{\text{$\nearrow$}} & &
            \gate{-1}
            \gategroup[6,steps=2,style={dotted,rounded corners, inner xsep=0.5pt},background,label style={label position=below,anchor=north,yshift=-0.2cm}]{\text{$\nwarrow$}} & &
            \gate{+1}
            \gategroup[6,steps=2,style={dotted,rounded corners, inner xsep=0.5pt},background,label style={label position=below,anchor=north,yshift=-0.2cm}]{\text{$\searrow$}} & &
            \gate{-1}
            \gategroup[6,steps=2,style={dotted,rounded corners, inner xsep=0.5pt},background,label style={label position=below,anchor=north,yshift=-0.2cm}]{\text{$\swarrow$}} & &
            \\
            \lstick{$\ket{q_{\text{grid}}}_y$} & & \gate{+1} & & \gate{+1} & & \gate{-1} & & \gate{-1} &
            \\
            \lstick{$\ket{q_{\text{dir},1}}_x$} & \octrl{-2} & \octrl{-1} & \ctrl{-2} & \ctrl{-1} & \octrl{-2} & \octrl{-1} & \ctrl{-2} & \ctrl{-1} & \\
            \lstick{$\ket{q_{\text{dir},2}}_x$} & \ctrl{-1} & \ctrl{-1} & \ctrl{-1} & \ctrl{-1} & \ctrl{-1} & \ctrl{-1} & \ctrl{-1} & \ctrl{-1} & \\
            \lstick{$\ket{q_{\text{dir},3}}_y$} & \octrl{-1} & \octrl{-1} & \octrl{-1} & \octrl{-1} & \ctrl{-1} & \ctrl{-1} & \ctrl{-1} & \ctrl{-1} & \\
            \lstick{$\ket{q_{\text{dir},4}}_y$} & \ctrl{-1} & \ctrl{-1} & \ctrl{-1} & \ctrl{-1} & \ctrl{-1} & \ctrl{-1} & \ctrl{-1} & \ctrl{-1} &
        \end{quantikz}
        \caption{D2Q9 diagonal streaming}
        \label{fig:D2Q9_diagonal_streaming_circ}
    \end{subfigure}

    \caption{Implementation of the D2Q9 streaming operations in different directions as a combination of $+1$ and $-1$ operations from figure \ref{fig:+1_and_-1_operations} with corresponding controls on the direction qubits as in table \ref{tab:D2Q9_arrangement}.}
    \label{fig:D2Q9_streaming_circ}
\end{figure}

For the D2Q9 scheme, additional up and down streaming operations are performed on $y$-direction grid qubits with corresponding control states on the direction qubits to select the correct subspace like arranged in table \ref{tab:D2Q9_arrangement}. The diagonal streaming directions are performed by sequentially performing a left or right with an up or down operations on the same direction subspace, so with similar control states on the direction register. The quantum circuits for the streaming of the directions in D2Q9 are shown in figure \ref{fig:D2Q9_streaming_circ}.

\subsubsection{Updating macroscopic variables}\label{sec:macro}

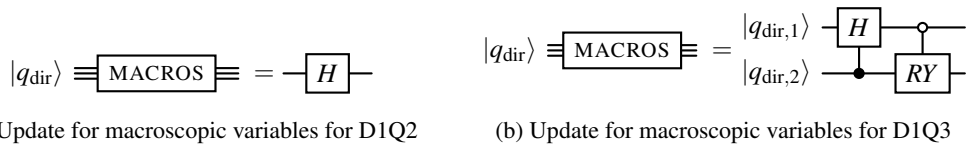
\begin{figure}[!htbp]
    \centering
    \begin{subfigure}{0.49\textwidth}
        \centering
        \begin{quantikz}[wire types={b}, classical gap=0.07cm, column sep=0.3cm]
           \lstick{$\ket{q_\text{dir}}$} & \gate{\text{\sc macros}} &
        \end{quantikz}
        $=$\begin{quantikz}[wire types={q}, classical gap=0.07cm, column sep=0.3cm]
            & \gate{H} &
        \end{quantikz}
        \caption{Update for macroscopic variables for D1Q2}
        \label{fig:D1Q2_macro_circ}
    \end{subfigure}
    \hspace*{-1cm}   
    \begin{subfigure}{0.49\textwidth}
        \centering
        \begin{quantikz}[wire types={b}, classical gap=0.07cm, column sep=0.2cm]
           \lstick{$\ket{q_\text{dir}}$} & \gate{\text{\sc macros}} &
        \end{quantikz}
        $=$\begin{quantikz}[wire types={q,q}, classical gap=0.07cm, row sep=0.1cm, column sep=0.2cm]
           \lstick{$\ket{q_{\text{dir},1}}$} & \gate{H} & \octrl{1} & \\
           \lstick{$\ket{q_{\text{dir},2}}$} & \ctrl{-1} &\gate{RY} &
        \end{quantikz}
        \caption{Update for macroscopic variables for D1Q3}
        \label{fig:D1Q3_macro_circ}
    \end{subfigure}

    \caption{Update of the macroscopic variables as the sum of all distributions into the first subspace for D1Q2 (fig. \ref{fig:D1Q2_macro_circ}) and D1Q3 (fig. \ref{fig:D1Q3_macro_circ}). }
    \label{fig:D1Q2_D1Q3_macro_circ}
\end{figure}

The summation of the distribution functions in the different subspaces is basically done by reversing the steps of distributing the distribution functions along the subspaces as is done in the collision steps, so by collecting instead of distributing subspaces. Assuming a total statevector of dimension $2N$, which can be subdivided into two distributions of dimension $N$, such that $\ket{\Psi} = (\Psi_1^1, \hdots, \Psi_1^N, \Psi_2^1, \hdots, \Psi_2^N)^T$, this can be represented by the two distributions $\ket{\Psi_1}, \ket{\Psi_2}$ for different states of the last, the ancilla, qubits:
\begin{equation}
    \ket{\Psi} = \ket{0}\otimes\ket{\Psi_1} + \ket{1}\otimes\ket{\Psi_2} \equiv (\Psi_1, \Psi_2)^T.
\end{equation}
The Hadamard transformation $H$ can distribute from a distribution in one subspace to another subspace with zero probability $(\Phi, 0)^T$ to two equal parts in both subspaces $(\Phi_{1/2}, \Phi_{1/2})^T$, while having to conserve the norm of the vector.
The Hadamard transformation operation $H$ is its self inverse, so while distributing from one subspace to two subspaces, it also reverses the operation and collects information back into the first subspace $(\Phi, 0)^T \overset{H}{\longleftrightarrow} (\Phi_{1/2}, \Phi_{1/2})^T$ when applied again. The Hadamard transformation calculates the equally weighted sum in the first subspace, while at the same time, the difference of the distributions is kept in the second subspace
\begin{equation}\label{eq:sum_by_H_gate}
    H\ket{a}\otimes\ket{\Psi}
    =
    H\ket{0}\otimes\ket{\Psi_1} +
    H\ket{1}\otimes\ket{\Psi_2}
    =
    \sqrt{\frac{1}{2}}\left(
        \ket{0}\otimes\ket{\Psi_1 + \Psi_2} +
        \ket{1}\otimes\ket{\Psi_1 - \Psi_2}
    \right),
\end{equation}
where we notice, that we calculate the sum in the first subsection up to a normalization factor of $\sqrt{1/2}$, which will account for another loss of signal, as also further taken into account in section \ref{sec:decay}.\\
The Hadamard transformation as used here calculates the equally weighted sum and difference up to a normalization of the two distributions. The same holds in general for the $RY$ gate with corresponding choice if $\theta$, which we understand as mixing in between the subspaces as has been explained in section \ref{sec:collision}.
Note that in order to maintain the sum and the difference in the corresponding subspaces $\ket{0}$ and $\ket{1}$, we need to choose $\theta$ negative, such that the cosine remains the same but the sine flips its sign and the difference shifts to the bottom line in the $RY$ matrix
\begin{equation}
    RY(\theta) \equiv
    \begin{pNiceArray}{>{\strut}rr}[margin,extra-margin = 1pt] 
        \cos(\theta / 2) & -\sin(\theta / 2) \\
        \sin(\theta / 2) & \cos(\theta / 2)
    \end{pNiceArray}
    =
    \begin{pNiceArray}{>{\strut}rr}[margin,extra-margin = 1pt] 
        \cos(-\theta / 2) & \sin(-\theta / 2) \\
        -\sin(-\theta / 2) & \cos(-\theta / 2)
    \end{pNiceArray}.
\end{equation}
The result of the difference state is of no further interest. With choosing different angles of $\theta$, the sum can be calculated with different weights. It is important to avoid including zero subspaces that do not contain any distribution function because this would lead to a significant additional decay of the solution. For the sequential summation of different subspaces we have to account for different weightings in each summation step, due to the general norm preservation of a quantum state. The weighted summation of two subspaces $\ket{\Psi_1}$ and $\ket{\Psi_2}$ using the $RY$-gates results in
\begin{equation}\label{eq:sum_by_RY_gate}
    RY(\theta)\ket{a}\otimes\ket{\Psi}
    =
    \ket{0}\otimes\ket{\cos(-\theta / 2)\Psi_1 + \sin(-\theta / 2)\Psi_2}
    +
    \ket{1}\otimes\ket{\cos(-\theta / 2)\Psi_2 - \sin(-\theta / 2)\Psi_1}.
\end{equation}
For the \textbf{D1Q2} scheme, one equally weighted sum is needed. This is done with the Hadamard gate as demonstrated in equation \eqref{eq:sum_by_H_gate}. The circuit is shown in figure \ref{fig:D1Q2_macro_circ}. \\
For the \textbf{D1Q3} scheme, at first, only the third and fourth subspaces, so $f_{\rightarrow}$ and $f_{\leftarrow}$ in $\Psi = (f_\text{rest}, 0, f_{\rightarrow}, f_{\leftarrow})^T$, are summed equally weighted by a Hadamard gate, which leaves its sum in the third subspace. Afterwards, a $RY$ gate is used to sum the first and third subspace where the sum results in the first subspace. We need the final sum to be an equally weighted sum of the first, third and fourth subspace, so $f_\text{rest}+f_{\rightarrow}+f_{\leftarrow}$, meaning a weighting of the statevector with $(1,0,1,1)^T$. Due to normalization, we need the weights to be $(\sqrt{1/3},\ 0, \sqrt{1/3}, \sqrt{1/3})$, which is fine as long as the weights remain equal. To apply the gates as summation operations on these isolated subspaces, controls need to placed accordingly. The quantum circuit with corresponding controls are shown in figure \ref{fig:D1Q3_macro_circ}. For these two controlled operations, the total summation operation on the two direction subspaces is:
\begin{align}
    RY(\theta)\bigg \vert_{\ket{q_{\text{dir},1}}=\ket{0}}
    H\bigg \vert_{\ket{q_{\text{dir},2}}=\ket{1}} \Psi
    &=
    \begin{pNiceArray}{rrrr}
       \cos\left(-\frac{\theta}{2}\right) & 0 & \sin\left(-\frac{\theta}{2}\right) & 0 \\
       0 & 0 & 0 & 0 \\
      -\sin\left(-\frac{\theta}{2}\right) & 0 & \cos\left(-\frac{\theta}{2}\right) & 0 \\
       0 & 0 & 0 & 0
    \end{pNiceArray}
    \begin{pNiceArray}{rrrr}
        1 & 0 & 0 & 0 \\
        0 & 1 & 0 & 0 \\
        0 & 0 & \sqrt{1/2} & \sqrt{1/2} \\
        0 & 0 & \sqrt{1/2} & -\sqrt{1/2}
    \end{pNiceArray}
    \begin{pmatrix}
        f_\text{rest} \\
        0  \\
        f_{\rightarrow} \\
        f_{\leftarrow}
    \end{pmatrix}
    \\
    &=
    \begin{pNiceArray}{cccc}
        \cos\left(-\frac{\theta}{2}\right) & 0 & \sin\left(-\frac{\theta}{2}\right)\sqrt{\frac{1}{2}} & \sin\left(-\frac{\theta}{2}\right)\sqrt{\frac{1}{2}} \\
        \cdot & \cdot & \cdot & \cdot  \\
        \cdot & \cdot & \cdot & \cdot  \\
        \cdot & \cdot & \cdot & \cdot
    \end{pNiceArray}
    \begin{pmatrix}
        f_\text{rest} \\
        0  \\
        f_{\rightarrow} \\
        f_{\leftarrow}
    \end{pmatrix},
\end{align}
which will fulfill our demand of an equally weighted summation $(\sqrt{1/3}f_\text{rest}+0+ \sqrt{1/3}f_{\rightarrow}+ \sqrt{1/3}f_{\leftarrow}, \ \cdot, \ \cdot, \ \cdot)^T$ by the condition:
\begin{align}
    & \cos\left(-\frac{\theta}{2}\right) \overset{!}{=} \sqrt{\frac{1}{3}}
    \hspace{0.2cm} \land \hspace{0.2cm}
    \sin\left(-\frac{\theta}{2}\right)\sqrt{\frac{1}{2}} \overset{!}{=} \sqrt{\frac{1}{3}}
    \\
    & \Longleftrightarrow \theta = - 2 \arccos\left(\sqrt{\frac{1}{3}}\right) = - 2 \arcsin\left(\sqrt{\frac{2}{3}}\right) \label{eq:theta_macro}.
\end{align}
So for the calculation of the D1Q3 macroscopic quantities with its result in the first subspace, a Hadamard gate on the first direction qubit with a control on the second direction qubit being in state $\ket{1}$ and a $RY$ rotation gate with $\theta = -\arccos(\sqrt{1/3})$ on the second direction qubit with control on the first direction qubit being in state $\ket{0}$ is used, as shown in figure \ref{fig:D1Q3_macro_circ}.\\
To calculate the macroscopic values in the \textbf{D2Q9} scheme, the summation is done in a similar way with the same summation weights as in the D1Q3 scheme, which is shown in the circuit in figure \ref{fig:D2Q9_macro_circ} with the summation order shown in table \ref{tab:D2Q9_summation_f} in the appendix. For the D2Q9 scheme, the summation is done in $x$- and $y$-direction, respectively, by the same operations as in the D1Q3 summation step. First, the $x$-direction summation operations perform the summations of the directions $\ket{\text{rest}}, \ket{\rightarrow}, \ket{\leftarrow}$ into $\ket{\text{rest}}$ directions, the $\ket{\uparrow}, \ket{\nwarrow}, \ket{\nearrow}$ direction
into $\ket{\uparrow}$ direction and $\ket{\downarrow}, \ket{\swarrow}, \ket{\searrow}$ directions
into $\ket{\downarrow}$ direction with the Hadamard gate $H$ and rotation gate $RY$ as in D1Q3. After that, the final summations of $\ket{\text{rest}}, \ket{\uparrow}, \ket{\downarrow}$ into$\ket{\text{rest}}$ are done similarly but on the $y$-qubits.

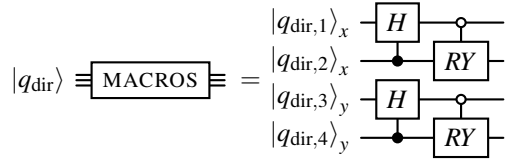
\begin{figure}[!htbp]
    \centering
    \begin{quantikz}[wire types={b}, classical gap=0.07cm, column sep=0.2cm]
        \lstick{$\ket{q_\text{dir}}$} & \gate{\text{\sc macros}} &
    \end{quantikz}
    $=$\begin{quantikz}[wire types={q,q,q,q}, classical gap=0.07cm, row sep=0.0cm, column sep=0.2cm]
       \lstick{$\ket{q_{\text{dir},1}}_x$} & \gate{H} & \octrl{1} & \\
       \lstick{$\ket{q_{\text{dir},2}}_x$} & \ctrl{-1} &\gate{RY} & \\
       \lstick{$\ket{q_{\text{dir},3}}_y$} & \gate{H} & \octrl{1} & \\
       \lstick{$\ket{q_{\text{dir},4}}_y$} & \ctrl{-1} &\gate{RY} &
   \end{quantikz}
    \caption{Update of the macroscopic variables as the sum of all distributions into the first subspace for the D2Q9 scheme using $\theta$ from equation \eqref{eq:theta_macro} for the $RY$ gates, following the order schematically shown in table \ref{tab:D2Q9_summation_f}.  }
    \label{fig:D2Q9_macro_circ}
\end{figure}

\subsubsection{Re-prepare state for next time step}\label{sec:re-prep}

\begin{figure}[!htbp]
    \centering
        \begin{quantikz}[wire types={b,b}, classical gap=0.07cm, column sep=0.3cm]
           \lstick{$\ket{q_\text{dir}}$} & \gate[2]{\text{\sc re-prep}} & \\
           \lstick{$\ket{q_\text{time}}$} &  &
        \end{quantikz}
        $=$\begin{quantikz}[wire types={b,b}, classical gap=0.07cm, column sep=0.3cm]
           \lstick{$\ket{q_\text{dir}}$} & \gate[2]{+N_Q} & \gate[2]{f_1 \leftrightarrow f_{N_{Q+1}}} & \\
           \lstick{$\ket{q_\text{time}}$} &  & &
       \end{quantikz}
        $=$\begin{quantikz}[wire types={b,b}, classical gap=0.07cm, column sep=0.3cm]
           \lstick{$\ket{q_\text{dir}}$} &  & \gate[2]{f_1 \leftrightarrow f_{N_{Q+1}}} & \\
           \lstick{$\ket{q_\text{time}}$} & \gate{+1} & &
       \end{quantikz}
       $=$\begin{quantikz}[wire types={q,q}, classical gap=0.07cm, column sep=0.3cm, row sep=0.3cm, align equals at = 4.5]
           \lstick{$\ket{q_{\text{dir},1}}$} &&  &  & & \octrl{2} & \\
            \wave&&&&&&  \\
           \lstick{$\ket{q_{\text{dir},N_Q}}$} & & &  & & \octrl{1} & \\
           \lstick{$\ket{q_{\text{time},1}}$} & \ctrl{1} & \ctrl{1} & \ctrl{1} & \gate{X} & \gate{X} & \\
           \lstick{$\ket{q_{\text{time},2}}$} & \ctrl{2} & \ctrl{2} & \gate{X} & & \octrl{-1} & \\
           \wave&&&&&&  \\
           \lstick{$\ket{q_{\text{time},N_{T-1}}}$} & \ctrl{1} & \gate{X} && & \octrl{-2} & \\
           \lstick{$\ket{q_{\text{time},N_{T}}}$} & \gate{X} &  && & \octrl{-1} &
       \end{quantikz}
    \caption{Quantum circuit of the \textsc{re-prep} step in figure \ref{fig:QLBM_circ_mult_steps} to prepare the state vector for the next time step in order to allow a fully quantum algorithm for all simulation time steps without the need of reinitialization or mid-circuit measurements. }
    \label{fig:re-prep_circ}
\end{figure}
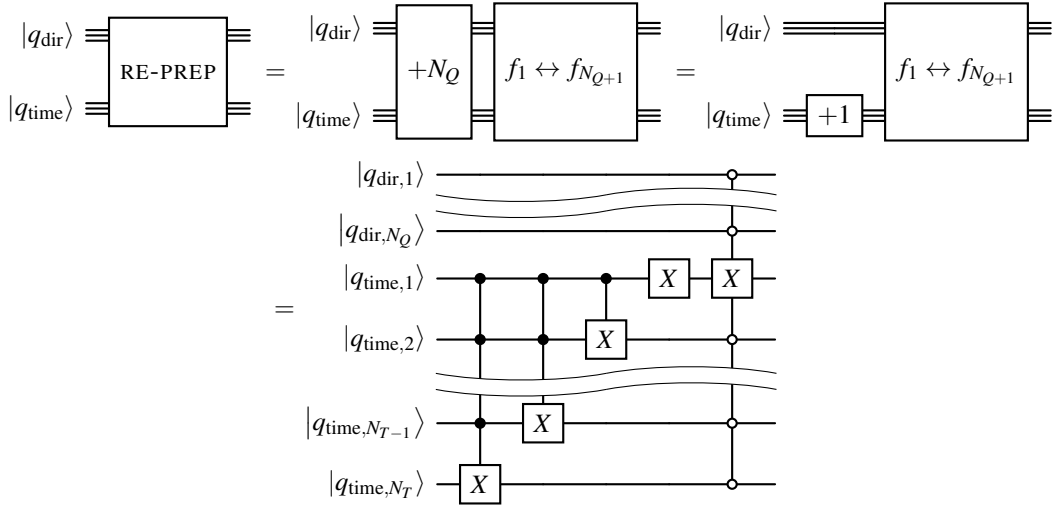

In order to create a fully quantum algorithm for multiple quantum lattice Boltzmann steps without reinitialization, the state vector has to be re-prepared to the state that contains the scalar grid information only in the first direction subspace while all other direction subspaces have to have zero state amplitudes, i.e. $(f_1,\ f_2,\ \hdots,\ f_{N_Q})^T = (\Phi,\ 0,\ \hdots,\ 0)^T$ (cf. section \ref{sec:enc_and_init_state}).
As we see in equation \eqref{eq:sum_by_H_gate}, with the summation of subspaces also comes a subtraction result. So after the calculation of the macroscopic quantities, there is the sum in the first subspace but generally also left over information in all other used direction subspaces. But in order to make the algorithm work for multiple time steps with collision, streaming and calculation of the macroscopic quantities, the state requires to have only the grid results in the first velocity direction subspace and all other velocity direction subspaces need to have zero probability amplitude, as stated in equation \eqref{eq:state_init}.
So a way is needed to set these probabilities to zero, without destroying the system quantum state. \\
The idea is to introduce additional qubits as part of the time qubit register, which is shown in figure \ref{fig:QLBM_circ_mult_steps}. The first additional time qubit doubles the number of direction subspaces, so for $N_Q$ direction subspaces, the statevector expands from $(f_1,\ f_2,\ \hdots,\ f_{N_Q})^T$ to $(f_1,\ f_2,\ \hdots,\ f_{N_Q},\ 0,\ \hdots,\ 0)^T$. This statevector now contains $N_Q$ additional subspaces with states of probability zero, because the additional time qubit, as long as not operated on, remains in state $\ket{0}$, so in the state of the first $N_Q$ direction subspaces and never takes $\ket{1}$ in the second $N_Q$ direction subspaces. The key part is to shift all $2N_Q$ periodically and switching the $N_Q+1$ subspace with the first subspace again:
\begin{equation}\label{eq:re-prep_state}
    \textsc{re-prep:}
    \begin{pmatrix}
        f_1\\
        f_2\\
        \vdots\\
        f_{N_Q}\\
        0\\
        0\\
        \vdots\\
        0
    \end{pmatrix}
    \longrightarrow
    \begin{pmatrix}
        0\\
        0\\
        \vdots\\
        0\\
        f_1\\
        f_2\\
        \vdots\\
        f_{N_Q}
    \end{pmatrix}
    \longrightarrow
    \begin{pmatrix}
        f_1\\
        0\\
        \vdots\\
        0\\
        0\\
        f_2\\
        \vdots\\
        f_{N_Q}
    \end{pmatrix},
\end{equation}
where after the collision, streaming and macroscopic calculation steps, the first subspace contains the gird $f_1 \sim \Phi$ up to a known normalization factor. The norm of $f_1$ will decrease in every time step due to the left over results of the subtraction in the other subspaces. This decay is further discussed in section \ref{sec:decay}.
Since the collision, streaming and macroscopic calculation steps only act within a group of every $N_Q$ subspaces, the state after the re-preparation at the end of equation \eqref{eq:re-prep_state} can be used for the next time step. For multiple time steps, further time qubits are added, where each additional time qubits doubles the number of zero-probability-subspaces. Therefore, the number of time qubits needed scales logarithmically with the number of simulated time steps $T$, so $\# q_\text{time} \sim \log(T)$, meaning that $\# q_\text{time}$ time qubits can simulate up to $T = 2^{\# q_\text{time}}$ time steps. The periodic shifting $+N_Q$ of the entries of state vector can be realized in a similar way as the streaming operation. The quantum circuit is shown in figure \ref{fig:re-prep_circ}.\\
For some specific lattice Boltzmann stencils, the \textsc{re-prep} circuit as shown in figure \ref{fig:re-prep_circ} can be simplified.
For the D1Q2 and D1Q3 stencil a simplification of the \textsc{re-prep} circuit is shown in the appendix in section \ref{sec:re-prep_D1Q2_and_D1Q3}.

\subsubsection{Decay of quantum state amplitude of solution per time step}\label{sec:decay}

The updated grid values for the next time step in the first subspace $f_1^{t+1} = \Phi^{t+1}$ undergoes a certain decay of its coefficient amplitude at two points in the algorithm: the collision operation and the calculation of the macroscopic variables. The first decay comes from the collision operation, that distributes the full grid in the first subspace $f_1$ to the velocity distributions $f_i$. In this step, only the normalized fraction $n_1$ of $\Phi$ is kept in the first subspace and the remaining signal $n'$ is moved to subspace $f_2$. Instead of keeping the full amount $k_1$ in subspace $f_1$, as would be dictated by the non-unitary collision operation in equation \eqref{eq:feq}, only the normalized amount of $n_1$ is kept. This normalization leads to the first signal decay in $f_1$ as determined in equation \eqref{eq:theta_coll_DxQx}
\begin{equation}\label{eq:gamme_coll}
    \gamma_{\text{coll},f_1} =\frac{1}{\sqrt{ \sum_{i=1}^Q k_j^2 }}.
\end{equation}
The second decay arises from the last summation step when calculating the macroscopic variables. As shown in the section \ref{sec:macro}, a summation of subspaces comes with a subtraction as well. While there is the sum of two subspaces in one subspace, there is the subtraction result in the other subspace. The resulting subtraction is some left over, unusable signal in the other subspaces. The last summation step is weighted with $\sqrt{1/Q}$ to take into account previous sequentially summations of velocity subspaces. So the decay due to the summation of the last subspaces that includes the $f_1$ subspace is
\begin{equation}\label{eq:gamme_macro}
    \gamma_{\text{macro},f_1} =\sqrt{\frac{1}{ Q }}.
\end{equation}
%
\newline
By determining only the decay of the macroscopic value $\Phi$, which is calculated in subspace $f_1$, only the collision decay and macroscopic variable calculation decay for $f_1$ need to be taken into account. Further decays due to the collision and summation in further velocity subspaces $f_{i>1}$ do not change the decay in the first subspace. Only when $f_1$ is included in the operations $f_1$ decays further. This results in a total decay of
\begin{equation}\label{eq:decay}
    \gamma_{\text{tot},f_1} =\frac{1}{\sqrt{ \sum_{i=1}^Q k_j^2 }}\sqrt{\frac{1}{ Q }},
\end{equation}
which is the signal loss of the complex amplitude coefficients of the macroscopic grid variables in the first velocity subspace per time step. For the decay of probability amplitude, the square of the decay value $\gamma_{\text{tot},f_1}^2$ determines the signal loss. In order to reconstruct the grid after $T$ time steps, that is encoded in the complex amplitude coefficients, the magnitude of the complex amplitude coefficients need to be multiplied by the decay factor for $T$ time steps $\gamma_{\text{tot},f_1}^T$.

\newpage
\section{Verification}

To verify our proposed algorithm, we model the advection-diffusion process of a gaussian distributed concentration in one and two dimensions for multiple time steps without reinitialization. We use our quantum lattice Boltzmann method (QLBM) with a D1Q2, D1Q3 and D2Q9 stencil and compare it to classical lattice Boltzmann (LBM) results. All simulations are performed such that each velocity distribution function fully relaxes to its local equilibrium within one time step, i.e. $\Delta t = \Delta \tau = 1$. Further, a spacial discretization of $\Delta x = 1$ is used.
With different sets of lattice Boltzmann weights $w_i$ different diffusion constants $D$ are modeled ensuring conservation of moment equations at least up to order four (cf. section \ref{sec:model_D_by_wi}).
For the simulation of our quantum algorithm, the shot method of the \textit{Qiskit} package \cite{qiskit2024} from \textit{IBM} is used.

\subsection{1D Advection-Diffusion equation}

\begin{figure}[!htbp]
    \begin{subfigure}{0.49\textwidth}
        \includegraphics[width=\linewidth]{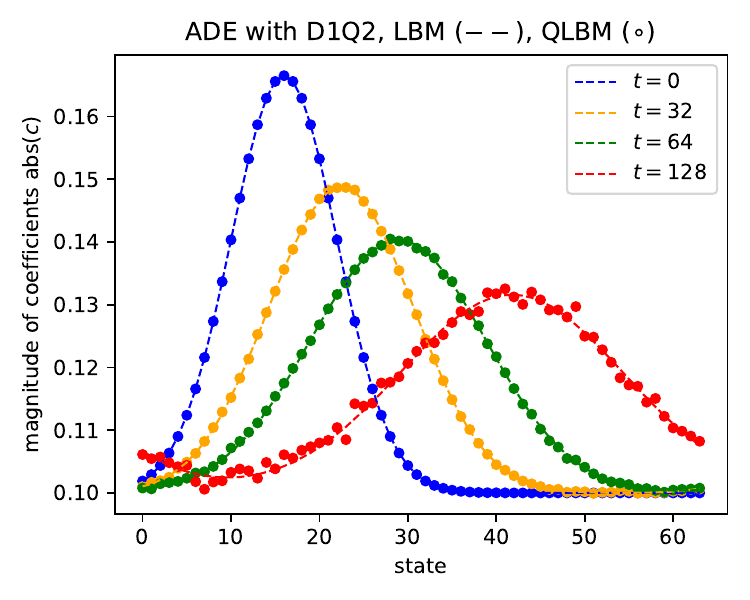}
        \caption{1D ADE with the D1Q2 scheme} \label{fig:QLBM_D1Q2}
    \end{subfigure}
    \hspace*{\fill}   
    \begin{subfigure}{0.49\textwidth}
        \includegraphics[width=\linewidth]{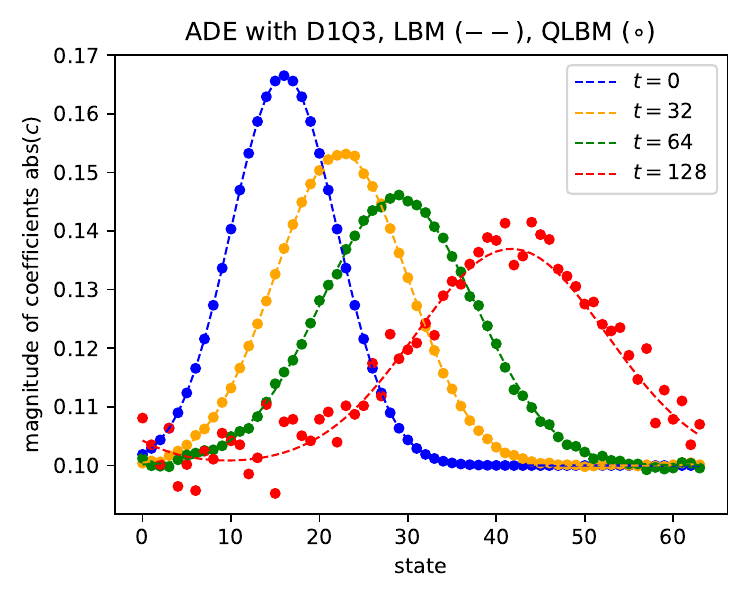}
        \caption{1D ADE with the D1Q3 scheme} \label{fig:QLBM_D1Q3}
    \end{subfigure}%

    \caption{Fully Quantum Lattice Boltzmann method without reinitialization for the 1D linear advection-diffusion equation solved with a D1Q2 scheme in figure \ref{fig:QLBM_D1Q2} and a D1Q3 scheme in figure \ref{fig:QLBM_D1Q3}. }
    \label{fig:QLBM_1D_ADE}
\end{figure}

The simulation of a one-dimensional gaussian hill following the ADE is done with a D1Q2 stencil and D1Q3. For both 1D test cases, a uniform velocity of $u = 0.2$ and periodic boundary conditions are used. The gaussian hills are initialized at position $x_0 = 16$ with a standard deviation of $\sigma = 6.0$ and a global offset of $o = 0.1$. The grid is discretized on $N = 64$ grid points by $\#q_\text{grid} = 6$ grid qubits. For a maximum simulation time of $t = 128$, a number of $\#q_\text{time} = 7$ qubits in the time qubit register are used. The D1Q2 stencil requires $\#q_\text{dir} = 1$ direction qubit for its two velocity directions, whereas the D1Q3 stencil requires $\#q_\text{dir} = 2$ direction qubits for its three velocity directions. The simulations are performed with $3.4 \cdot 10^7$ shots in total. The results are shown in figure \ref{fig:QLBM_D1Q2} and \ref{fig:QLBM_D1Q3}.\\
For the D1Q2 stencil, the standard LBM weights are used, i.e. [1/2, 1/2], which results in a squared speed of sound of $c_s^2 = 1$ and therefore a diffusion of $D = 1/2$ is modeled. This results in a decay per time step to $\gamma = 0.98$, which reduces the probability per time step to $\gamma^2 = 0.96$. \\
The simulation with the D1Q3 stencil uses the non-standard weight set of [1/3, 1/3, 1/3] which results in a squared speed of sound of $c_s^2 = 2/3$ and therefore a simulated diffusion of $D = 1/3 $. This setting results in a decay per time step to $\gamma = 0.97$, so a probability reduction to $\gamma^2 = 0.94$ per time step.\\
The results show that our QLBM is capable of reproducing the LBM results overall very accurately. For more time steps, the probability of measuring the correct states becomes less likely, which is due to the decay of the probability function in our algorithm. This results in an increasing noise of the solution for larger time steps. Performing the simulation with more shots in total reduces the noise, as will be discussed in more detail for the 2D test case in section \ref{sec:2D_ADE_results}.
\subsection{2D Advection-Diffusion equation}
\label{sec:2D_ADE_results}
\begin{figure}[!htbp]
    \begin{subfigure}{0.49\textwidth}
        \includegraphics[width=1.0\linewidth]{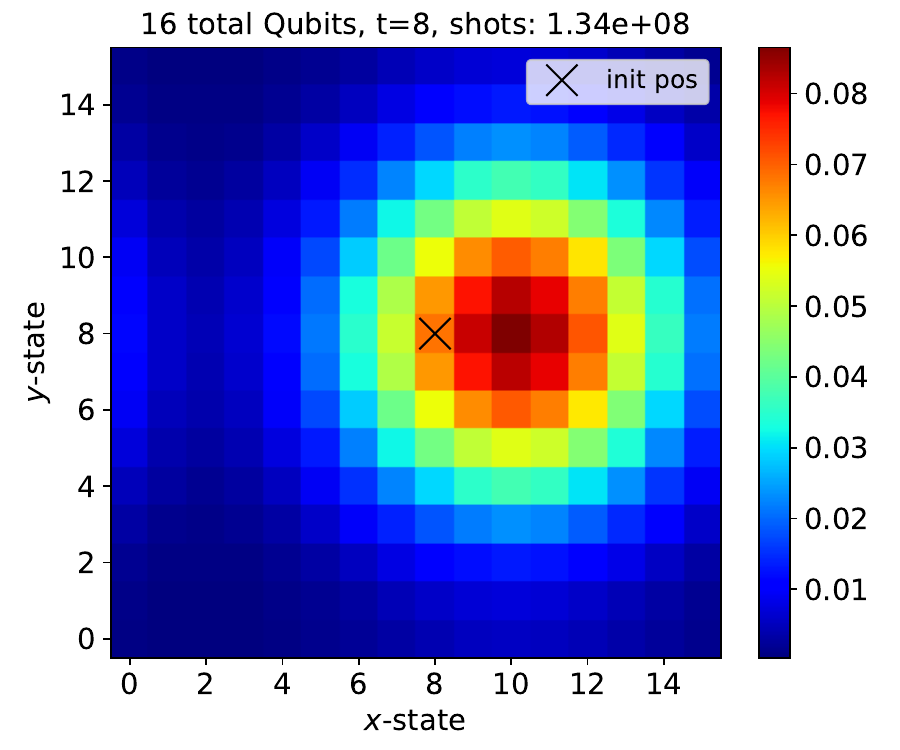}
        \caption{D2Q9. The flow velocity is $\mathbf{u} = (1/4, 0)^T$ in lattice units for $t = 8$ time steps. }
        \label{fig:QLBM_2D_ADE_t8}
    \end{subfigure}
    \hspace*{\fill}   
    \begin{subfigure}{0.49\textwidth}
        \includegraphics[width=1.0\linewidth]{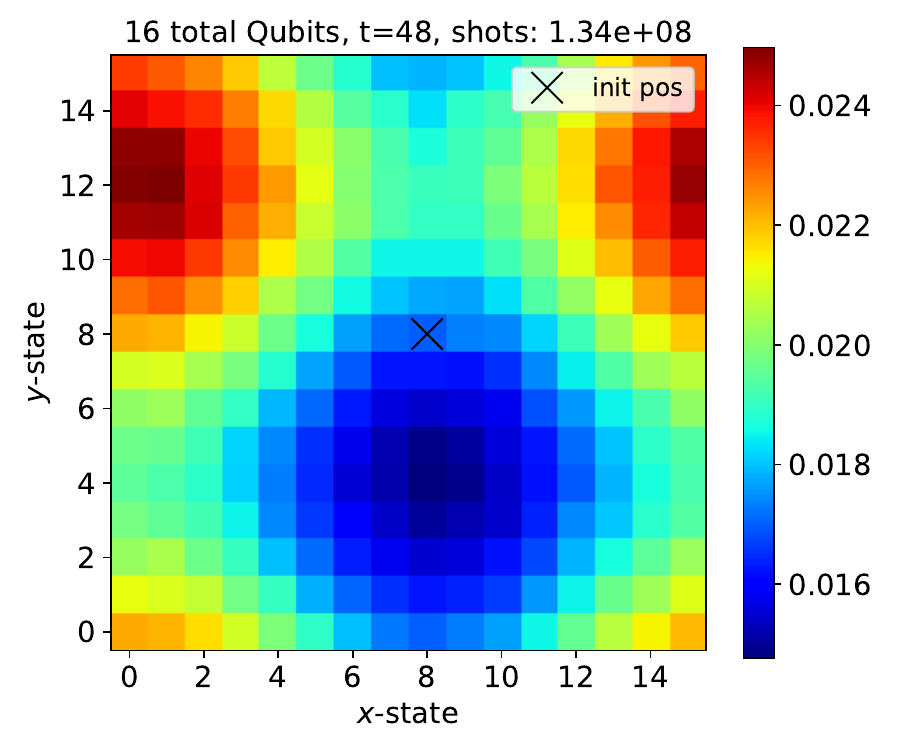}
        \caption{D2Q9. The flow velocity is $\mathbf{u} = (1/6, 1/12)^T$ in lattice units for $t = 48$ time steps. }
        \label{fig:QLBM_2D_ADE_t48}
    \end{subfigure}
    \caption{Fully Quantum Lattice Boltzmann method without reinitialization for the 2D linear advection-diffusion equation solved with a D2Q9 scheme with  periodic boundary conditions and a diffusion constant of $D = 1/3$. The cross ($\times$) indicates the initial position of the gaussian with a standard deviation of $\sigma = 2.0$. Two different times and velocities are simulated in figure \ref{fig:QLBM_2D_ADE_t8} and \ref{fig:QLBM_2D_ADE_t48}. }
    \label{fig:QLBM_2D_ADE}
\end{figure}

\begin{figure}[!htbp]
    \begin{subfigure}{0.49\textwidth}
        \includegraphics[width=1.0\linewidth]{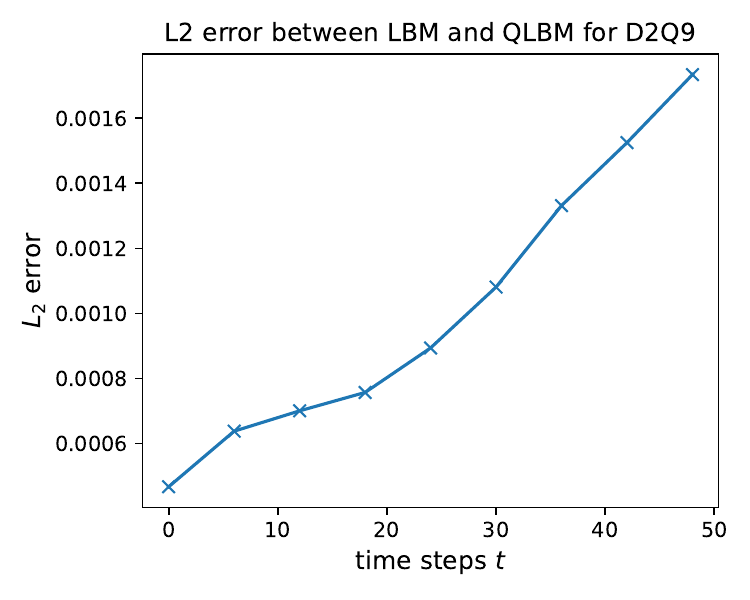}
        \caption{Evolution of the $L_2$-error deviation of QLBM to LBM method over time for the QLBM D2Q9 test case in fig. \ref{fig:QLBM_2D_ADE_t48}.}
        \label{fig:err_time}
    \end{subfigure}
    \hspace*{\fill}   
    \begin{subfigure}{0.49\textwidth}
        \includegraphics[width=1.0\linewidth]{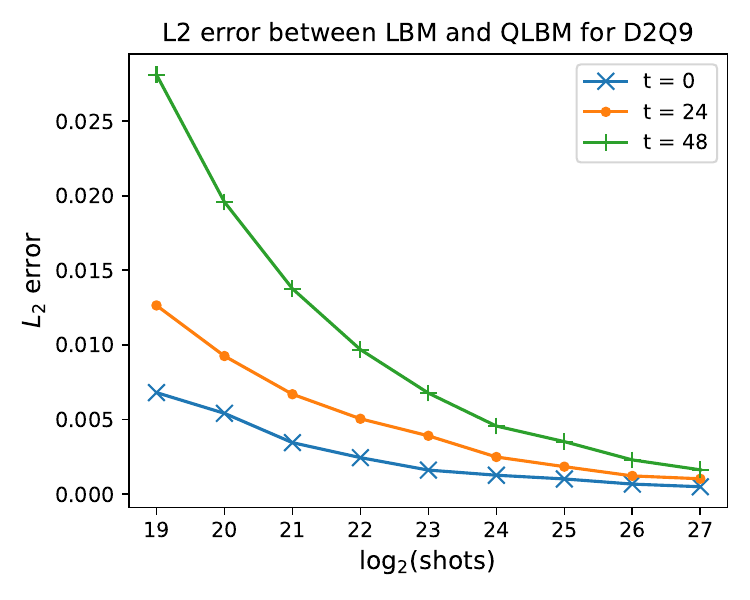}
        \caption{Convergence of the QLBM D2Q9 test case towards LBM for more shots at different simulation times. }
        \label{fig:err_shots}
    \end{subfigure}
    \caption{Comparison of QLBM and LBM method for a D2Q9 test case in figure \ref{fig:QLBM_2D_ADE_t48} with an initial gaussian distribution of standard deviation of size $\sigma = 2.0$ at a diffusion of $D = 1/3$ for a velocity of $\mathbf{u} = (1/6, 1/12)^T$. }
    \label{fig:err}
\end{figure}

To verify our algorithm for two-dimensions, we again propagate a gaussian hill, now in 2D, following the advection-diffusion equation using the D2Q9 stencil with periodic boundary conditions. The results for two different advection velocities are shown in figures \ref{fig:QLBM_2D_ADE_t8} and \ref{fig:QLBM_2D_ADE_t48}. \\
The gaussians are discretized on $16 \times 16$ nodes using  $\#q_\text{grid} = 4 + 4$ grid qubits for the $x$ and $y$ directions. The concentrations have a standard deviation of $\sigma = 2.0$ and no global offset. The D2Q9 stencil requires $Q = 9$ velocity directions, so $N_Q = 16$ velocity direction subspaces need to be generated, which is done using $\#q_\text{dir} = 4$ velocity direction qubits. With $\#q_\text{time} = 6$ time qubits, up to $T = 48$ time steps are simulated. The flow field is sampled at the end of the simulation with $1.3 \cdot 10^8$ shots.\\
Since the decay is smaller for smaller  difference of the values of the weights, we choose a weight set of equal weights. For the test cases, a set of $w_i = 1/9 \ \forall i \in [1,9] $ is used which results in the minimal decay of our solution. This set of weights simulates a squared speed of sound of $c_s = 2/3$ and thus a diffusion of $D = 1/3$ is simulated. The decay has a dependence on the advection velocity since it is part of the $k_i$ factors in the decay in equation \eqref{sec:decay}. So for the test case in figure \ref{fig:QLBM_2D_ADE_t8}, an advection velocity of $\mathbf{u} = (1/4, 0)^T$ is used, resulting in a decay of $\gamma = 0.96$ per time step, so a probability decay to $\gamma = 0.91$ per time step. For the test case in figure \ref{fig:QLBM_2D_ADE_t48}, the advection smaller velocity of $\mathbf{u} = (1/6, 1/12)^T$ results in a decay of $\gamma = 0.97$ per time step, so a probability decay to $\gamma = 0.95$ per time step. The results show very good agreement with the expectations, including deviations due to sampling noise. To verify the agreement with the classical LBM solutions, especially regarding expected sampling deviations, a more detailed look comparing QLBM with LBM is done: to quantify the difference the results by our QLBM compared to a classical LBM, we calculate the $L_2$ error as
\begin{equation}\label{eq:L2}
    L_2 = \sqrt{\sum_{i=1}^N (\Phi_\text{QLBM} - \Phi_\text{LBM})^2}.
\end{equation}
Due to an increasing decay of our solution for more time steps, we expect that for a fixed number of shots, fewer shots resolve our flow field in the correct subspace. Therefore, we expect increase deviation for an increasing number of simulated time steps between QLBM and LBM, since the flow field is effectively resolved with fewer shots. This is confirmed by calculating the $L_2$ error for more time steps, as is shown in figure \ref{fig:err_time}. Now this also means, that the error should decrease to zero for a fixed number of time steps with more sampling shots, meaning that the QLBM solution converges towards the LBM solution. This in fact can be verified by figure \ref{fig:err_shots}, which shows the convergence of QLBM solution towards the LBM solution for different simulation time lengths. This means that our QLBM algorithm can approximate the LBM solution arbitrarily close even for longer simulation times if the decay can be compensated with more total simulation shots.

\newpage
\section{Conclusion}

In this paper, an extension of the Quantum Lattice Boltzmann Method (QLBM) is proposed and verified, such that multiple time steps can be performed without the need of state measurement or reinitialization in between the time steps. This extension is valid for general lattice Boltzmann velocity stencils and tested on D1Q2, D1Q3 and D2Q9 stencils using the \textit{shot} methods of the \textit{Qiskit} simulation package. The algorithm is proposed and discussed in detail, giving the mathematical description as well as the quantum circuit diagrams. For the extended QLBM algorithm, we discuss our required initialization state, the collision and streaming step, calculation of macroscopic variables and a re-preparation step for the next time step, all as fully quantum algorithm blocks with corresponding quantum circuit gate diagrams.  \\
The extension is tested on a linear advection-diffusion equation (ADE) in one and two dimensions and compared to classical lattice Boltzmann (LBM) reference solutions. We show excellent agreement and a convergence of our QLBM to LBM for any desired accuracy. For very large, highly-resolved grids, a state extraction of the full grid may be infeasible. The main advantage of our algorithm is that there is no need to extract the full flow field at any time. While other algorithms requires measurements and state reinitialization of the full flow field, our algorithm can perform all time steps without any measurements or reinitialization. When only interested in surface integrals or scalar properties, our algorithm would allow to calculate these quantities without ever having to extract the flow field at any time at all. This overcomes the scaling issues of algorithms in the literature that require state extraction where the computational effort, given by the number of shots, scales with the grid resolution. Future work is dedicated to reduce the decay by such techniques like amplitude amplification and to tackle nonlinearities to be able to solve fluid flow equations like the \textit{Burgers} equations.

\subsection{Discussion of our algorithms advantages}
\label{sec:discussion}

Our goal is to investigate possible quantum algorithm approaches and improve these algorithms such that they may be suitable and usable for applications in Aerospace science. Since we are looking for algorithms that can deal with very large grids with extremely high resolution, it is important to find an algorithm where the computational cost scales efficiently with the grid resolution. These computational costs are essentially the number of shots and the number of gates required. An algorithm that requires to sample the fully resolved grid cannot fulfill these requirements. Therefore, we propose a method that can perform the full simulation without the need of state extraction at any time and still obtain the results that we are interested in. These can be mainly reduced quantities like lift, drag, or other body surface properties.\\
Our algorithm can perform multiple time steps more efficient in terms of the number of shots required than those which require state extraction in between every time step, although this efficiency holds only for a limited number of time steps. In terms of efficiency, one could think of combining our algorithm with state extraction algorithms, but for our purposes, we want to avoid a state extraction for the aforementioned reasons at all and only extract our reduced quantities. \\
A large computational advantage of our algorithm is that we can avoid having to reinitialize a very complex flow field in between the time steps, which generally will require a large number of gates to tune the statevector accordingly. For a fluid flow simulation around a body, we could basically initialize a uniform velocity field which is achieved with very few gates. \\
While our algorithm can be more efficient in the initialization and number of shots, we do not reduce the gate count compared to other QLBM algorithms. In fact, our algorithm without state extraction produces a very large circuit depth which may cause coherence time issues.

\subsection*{Acknowledgements}

This project was made possible by the DLR Quantum Computing Initiative and the Federal Ministry for Economic Affairs and Climate Action; \textit{qci.dlr.de/projects/toquaflics}.\\
Figures with Quantum circuits are visualized using the \textit{Quantikz} latex package \cite{kay2018tutorial}.

\section{Appendix}

\subsection{D2Q9 collision step distribution of velocity direction distribution functions}
\label{sec:appendix_coll_distr_D2Q9}

The distribution of the D2Q9 velocity distribution functions is easier to handle if not distributed directly in a sequential way. The order of the procedure of distributing the velocity distribution function into the corresponding direction subspaces by several $RY$-gate operations is shown in table \ref{tab:D2Q9_distributing_f}.

\begin{table}[!htbp]
\small
    \centering
    \caption{Distributing the D2Q9 velocity distribution functions along the velocity direction subspaces. The division of the four direction qubits is done such that the first two qubits form the $x$-directions and the last two qubits form the $y$-directions, so $\ket{q_4 q_3 q_2 q_1}_\text{dir} = \ket{q_4 q_3}_y\ket{q_2 q_1}_x$. The index of $\theta$ contains the state with decimal representation of the binary states.}
    \begin{tabular}{c|c|c|c|c}
         & $\ket{00}_x$ & $\ket{01}_x$ & $\ket{10}_x$ & $\ket{11}_x$ \\ \hline
        $\ket{00}_y$ & $f_1,\ \hdots, f_9$ & - & - & - \\ \hline
        $\ket{01}_y$ & - & - & - & -\\ \hline
        $\ket{10}_y$ & -  & - & - & - \\ \hline
        $\ket{11}_y$ & - & - & - & - \\
    \end{tabular}
    {\large $\xrightarrow{\theta_{\ket{0}\rightarrow\ket{2}}}$}
    \begin{tabular}{c|c|c|c|c}
         & $\ket{00}_x$ & $\ket{01}_x$ & $\ket{10}_x$ & $\ket{11}_x$ \\ \hline
        $\ket{00}_y$ & ${f_1, f_4, f_5}$ & - & ${f_2, f_3, f_6, f_7, f_8, f_9}$ & - \\ \hline
        $\ket{01}_y$ & - & - & - & -\\ \hline
        $\ket{10}_y$ & -  & - & - & - \\ \hline
        $\ket{11}_y$ & - & - & - & - \\
    \end{tabular}
    {\large $\xrightarrow{\theta_{\ket{2}\rightarrow\ket{3}}}$}
    \begin{tabular}{c|c|c|c|c}
         & $\ket{00}_x$ & $\ket{01}_x$ & $\ket{10}_x$ & $\ket{11}_x$ \\ \hline
        $\ket{00}_y$ & ${f_1, f_4, f_5}$ & - & ${f_2, f_6, f_8}$ & ${f_3, f_7, f_9}$ \\ \hline
        $\ket{01}_y$ & - & - & - & - \\ \hline
        $\ket{10}_y$ & -  & - & - & - \\ \hline
        $\ket{11}_y$ & - & - & - & - \\
    \end{tabular}
    {\large $\xrightarrow{\theta_{\ket{0}\rightarrow\ket{8}}}$}
    \begin{tabular}{c|c|c|c|c}
         & $\ket{00}_x$ & $\ket{01}_x$ & $\ket{10}_x$ & $\ket{11}_x$ \\ \hline
        $\ket{00}_y$ & ${f_1}$ & - & ${f_2, f_6, f_8}$ & ${f_3, f_7, f_9}$ \\ \hline
        $\ket{01}_y$ & - & - & - & - \\ \hline
        $\ket{10}_y$ & ${f_4, f_5}$  & - & - & - \\ \hline
        $\ket{11}_y$ & - & - & - & -  \\
    \end{tabular}
    {\large $\xrightarrow{\theta_{\ket{8}\rightarrow\ket{12}}}$}
    \begin{tabular}{c|c|c|c|c}
         & $\ket{00}_x$ & $\ket{01}_x$ & $\ket{10}_x$ & $\ket{11}_x$ \\ \hline
        $\ket{00}_y$ & ${f_1}$ & - & ${f_2, f_6, f_8}$ & ${f_3, f_7, f_9}$ \\ \hline
        $\ket{01}_y$ & - & - & - & - \\ \hline
        $\ket{10}_y$ & ${f_4}$  & - & - & - \\ \hline
        $\ket{11}_y$ & $f_5$ & - & - & -  \\
    \end{tabular}
    {\large $\xrightarrow{\theta_{\ket{2}\rightarrow\ket{10}}}$}
    \begin{tabular}{c|c|c|c|c}
         & $\ket{00}_x$ & $\ket{01}_x$ & $\ket{10}_x$ & $\ket{11}_x$ \\ \hline
        $\ket{00}_y$ & ${f_1}$ & - & ${f_2}$ & ${f_3, f_7, f_9}$ \\ \hline
        $\ket{01}_y$ & - & - & - & - \\ \hline
        $\ket{10}_y$ & ${f_4}$  & - & $f_6, f_8$ & - \\ \hline
        $\ket{11}_y$ & $f_5$ & - & - & -  \\
    \end{tabular}
    \newline
    {\large $\xrightarrow{\theta_{\ket{3}\rightarrow\ket{11}}}$}
    \begin{tabular}{c|c|c|c|c}
         & $\ket{00}_x$ & $\ket{01}_x$ & $\ket{10}_x$ & $\ket{11}_x$ \\ \hline
        $\ket{00}_y$ & ${f_1}$ & - & ${f_2}$ & ${f_3}$ \\ \hline
        $\ket{01}_y$ & - & - & - & - \\ \hline
        $\ket{10}_y$ & ${f_4}$  & - & $f_6, f_8$ & $f_7, f_9$ \\ \hline
        $\ket{11}_y$ & $f_5$ & - & - & -  \\
    \end{tabular}
    {\large $\xrightarrow{\theta_{\ket{10}\rightarrow\ket{14}}}$}
    \begin{tabular}{c|c|c|c|c}
         & $\ket{00}_x$ & $\ket{01}_x$ & $\ket{10}_x$ & $\ket{11}_x$ \\ \hline
        $\ket{00}_y$ & ${f_1}$ & - & ${f_2}$ & ${f_3}$ \\ \hline
        $\ket{01}_y$ & - & - & - & - \\ \hline
        $\ket{10}_y$ & ${f_4}$  & - & $f_6$ & $f_7, f_9$ \\ \hline
        $\ket{11}_y$ & $f_5$ & - & $f_8$ & -  \\
    \end{tabular}
    \newline
    {\large $\xrightarrow{\theta_{\ket{11}\rightarrow\ket{15}}}$}
    \begin{tabular}{c|c|c|c|c}
         & $\ket{00}_x$ & $\ket{01}_x$ & $\ket{10}_x$ & $\ket{11}_x$ \\ \hline
        $\ket{00}_y$ & ${f_1}$ & - & ${f_2}$ & ${f_3}$ \\ \hline
        $\ket{01}_y$ & - & - & - & - \\ \hline
        $\ket{10}_y$ & ${f_4}$  & - & $f_6$ & $f_7$ \\ \hline
        $\ket{11}_y$ & $f_5$ & - & $f_8$ & $f_9$  \\
    \end{tabular}      %
    \label{tab:D2Q9_distributing_f}
\end{table}

\newpage
\subsection{D2Q9 summation step of velocity direction distribution functions}
\label{sec:appendix_macro_D2Q9}

The order of the summation of the D2Q9 velocity distribution functions is shown in table \ref{tab:D2Q9_summation_f} of the corresponding circuit in figure \ref{fig:D2Q9_macro_circ}.

\begin{table}[!htbp]
\small
\centering
\caption{Summation order of the D2Q9 velocity distribution functions first in $x$-direction and then in $y$-direction.}
{\color{white} \large $\xrightarrow{H_{\ket{11}_x\rightarrow\ket{10}_x}}$}
    \begin{tabular}{c|c|c|c|c}
         & $\ket{00}_x$ & $\ket{01}_x$ & $\ket{10}_x$ & $\ket{11}_x$ \\ \hline
        $\ket{00}_y$ & ${f_1}$ & - & ${f_2}$ & ${f_3}$ \\ \hline
        $\ket{01}_y$ & - & - & - & - \\ \hline
        $\ket{10}_y$ & ${f_4}$  & - & $f_6$ & $f_7$ \\ \hline
        $\ket{11}_y$ & $f_5$ & - & $f_8$ & $f_9$  \\
    \end{tabular}
    {\large $\xrightarrow{H_{\ket{11}_x\rightarrow\ket{10}_x}}$}
    \begin{tabular}{c|c|c|c|c}
         & $\ket{00}_x$ & $\ket{01}_x$ & $\ket{10}_x$ & $\ket{11}_x$ \\ \hline
        $\ket{00}_y$ & ${f_1}$ & - & ${f_2, f_3}$ & - \\ \hline
        $\ket{01}_y$ & - & - & - & - \\ \hline
        $\ket{10}_y$ & ${f_4}$  & - & $f_6, f_7$ & - \\ \hline
        $\ket{11}_y$ & $f_5$ & - & $f_8, f_9$ & -  \\
    \end{tabular}
    \newline
    {\large $\xrightarrow{RY_{\ket{10}_x\rightarrow\ket{00}_x}}$}
    \begin{tabular}{c|c|c|c|c}
         & $\ket{00}_x$ & $\ket{01}_x$ & $\ket{10}_x$ & $\ket{11}_x$ \\ \hline
        $\ket{00}_y$ & $f_1, f_2, f_3$ & - & - & - \\ \hline
        $\ket{01}_y$ & - & - & - & -\\ \hline
        $\ket{10}_y$ & $f_4, f_6, f_7$  & - & - & - \\ \hline
        $\ket{11}_y$ & $f_5, f_8, f_9$ & - & - & - \\
    \end{tabular}
    {\large $\xrightarrow{H_{\ket{11}_y\rightarrow\ket{10}_y}}$}
    \begin{tabular}{c|c|c|c|c}
         & $\ket{00}_x$ & $\ket{01}_x$ & $\ket{10}_x$ & $\ket{11}_x$ \\ \hline
        $\ket{00}_y$ & $f_1, f_2, f_3$ & - & - & - \\ \hline
        $\ket{01}_y$ & - & - & - & -\\ \hline
        $\ket{10}_y$ & $f_4, f_6, f_7$  & - & - & - \\
        & $f_5, f_8, f_9$  &  &  &  \\ \hline
        $\ket{11}_y$ & - & - & - & - \\
    \end{tabular}
    \newline
    {\large $\xrightarrow{RY_{\ket{10}_y\rightarrow\ket{00}_y}}$}
    \begin{tabular}{c|c|c|c|c}
         & $\ket{00}_x$ & $\ket{01}_x$ & $\ket{10}_x$ & $\ket{11}_x$ \\ \hline
        $\ket{00}_y$ & $f_1,\ \hdots, f_9$ & - & - & - \\ \hline
        $\ket{01}_y$ & - & - & - & -\\ \hline
        $\ket{10}_y$ & -  & - & - & - \\ \hline
        $\ket{11}_y$ & - & - & - & - \\
    \end{tabular}
    \label{tab:D2Q9_summation_f}
\end{table}

\subsection{Re-prepare state for D1Q2 and D1Q3}
\label{sec:re-prep_D1Q2_and_D1Q3}

\begin{figure}[!htbp]
    \centering
        \begin{quantikz}[wire types={b,b}, classical gap=0.07cm, column sep=0.3cm]
           \lstick{$\ket{q_\text{dir}}$} & \gate[2]{\text{\sc re-prep}} & \\
           \lstick{$\ket{q_\text{time}}$} &  &
        \end{quantikz}
        $=$
        \begin{quantikz}[wire types={q,q,q,b}, classical gap=0.07cm, column sep=0.3cm, row sep=0.3cm, align equals at = 2.5]
            \lstick{$\ket{q_{\text{dir},1}}$} & &  \\
            \wave&&&&&&  \\
            \lstick{$\ket{q_{\text{dir},N_Q}}$} & \ctrl{1}& \\
            \lstick{$\ket{q_{\text{time}}}$} & \gate{+1} &
       \end{quantikz}
        $=$
        \begin{quantikz}[wire types={}, classical gap=0.07cm, column sep=0.3cm, row sep=0.3cm, align equals at = 4.5]
           \lstick{$\ket{q_{\text{dir},1}}$} &&  &  & &  \\
            \wave&&&&&&  \\
           \lstick{$\ket{q_{\text{dir},N_Q}}$} & \ctrl{1}&\ctrl{1} & \ctrl{1} & \ctrl{1}& \\
           \lstick{$\ket{q_{\text{time},1}}$} & \ctrl{1} & \ctrl{1} & \ctrl{1} & \gate{X} &  \\
           \lstick{$\ket{q_{\text{time},2}}$} & \ctrl{2} & \ctrl{2} & \gate{X} & &  \\
           \wave&&&&&&  \\
           \lstick{$\ket{q_{\text{time},N_{T-1}}}$} & \ctrl{1} & \gate{X} && & \\
           \lstick{$\ket{q_{\text{time},N_{T}}}$} & \gate{X} &  && &
       \end{quantikz}
    \caption{Simplified quantum circuit of \textsc{re-prep} step from figure \ref{fig:re-prep_circ} for the D1Q2 and D1Q3 state arrangements.}
    \label{fig:re-prep_circ_D1Q2_D1Q3}
\end{figure}
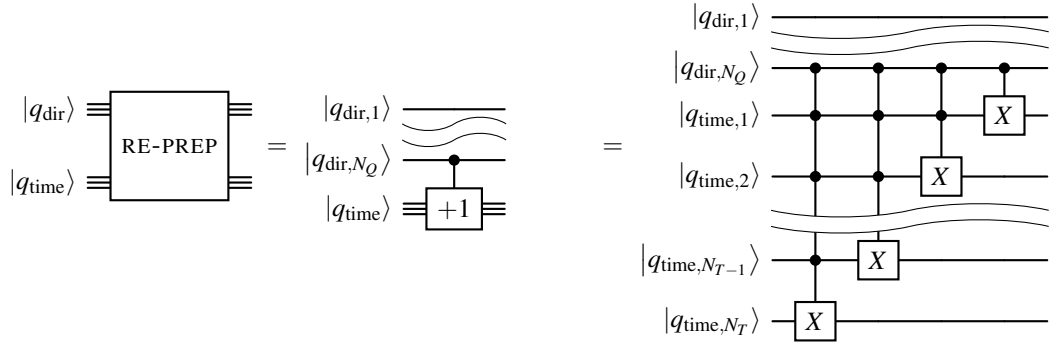

For the D1Q2 and D1Q3 stencil, the \textsc{re-prep} circuit shown in figure \ref{fig:re-prep_circ} can be simplified.
In these stencils, all the velocity subspaces but the first one are located in the second half of all subspaces, so $f_1$ is in subspace one and all velocity subspaces $f_i$ are in subspace with index $i>\frac{1}{2}N_Q$. This is shown in equations \eqref{eq:D1Q2_re-prep} and \eqref{eq:D1Q3_re-prep}, where the dashed line indicates the split into half of the velocity direction subspaces and the solid separation line the split to the additional subspaces due to the added time qubits.
This results in a \textsc{re-prep} step for the D1Q2 scheme of
\begin{equation}\label{eq:D1Q2_re-prep}
    \begin{pmatrix}
        \ket{0}_\text{time}\ket{0}_\text{dir}\\
        \hdashline
        \ket{0}_\text{time}\ket{1}_\text{dir}\\
        \hline
        \ket{1}_\text{time}\ket{0}_\text{dir}\\
        \ket{1}_\text{time}\ket{1}_\text{dir}
    \end{pmatrix}:
    \hspace{1cm}
    \begin{pmatrix}
        f_1\\
        \hdashline
        f_2\\
        \hline
        0 \\
        0
    \end{pmatrix}
    \xrightarrow{\text{D1Q2 \textsc{re-prep}}}
    \begin{pmatrix}
        f_1\\
        \hdashline
        0 \\
        \hline
        0 \\
        f_2
    \end{pmatrix}
\end{equation}
and for the D1Q3 scheme of
\begin{equation}\label{eq:D1Q3_re-prep}
    \begin{pmatrix}
        \ket{0}_\text{time}\ket{00}_\text{dir}\\
        \ket{0}_\text{time}\ket{01}_\text{dir}\\
        \hdashline
        \ket{0}_\text{time}\ket{10}_\text{dir}\\
        \ket{0}_\text{time}\ket{11}_\text{dir}\\
        \hline
        \ket{1}_\text{time}\ket{00}_\text{dir}\\
        \ket{1}_\text{time}\ket{01}_\text{dir}\\
        \ket{1}_\text{time}\ket{10}_\text{dir}\\
        \ket{1}_\text{time}\ket{11}_\text{dir}
    \end{pmatrix}:
    \hspace{1cm}
    \begin{pmatrix}
        f_1\\
        0\\
        \hdashline
        f_2\\
        f_3\\
        \hline
        0 \\
        0 \\
        0 \\
        0
    \end{pmatrix}
    \xrightarrow{\text{D1Q3 \textsc{re-prep}}}
    \begin{pmatrix}
        f_1\\
        0\\
        \hdashline
        0 \\
        0 \\
        \hline
        0 \\
        0 \\
        f_2\\
        f_3
    \end{pmatrix}    .
\end{equation}
In these arrangements it is sufficient to move only the second half of all velocity subspaces. With this arrangement of the velocity direction subspaces, this re-preparation can be achieved by a $+N_Q$ operation on all subspaces where the most significant qubit is in state $\ket{1}$ or equivalently a $+1$ operation on all time qubits conditioned on the most significant direction qubit to be in state $\ket{1}$. This circuit is shown in figure \ref{fig:re-prep_circ_D1Q2_D1Q3}.

\newpage
\addcontentsline{toc}{section}{References}
\printbibliography

\end{document}